\documentclass[prd,twocolumn,nofootinbib,superscriptaddress]{revtex4-2}
\usepackage{bm}
\usepackage{amsmath}
\usepackage{amssymb}
\usepackage{graphicx}
\usepackage{subfigure}
\usepackage{hyperref}
\usepackage{color}
\usepackage{comment}
\usepackage{ulem}
\usepackage{CJK}
\hypersetup{
	colorlinks=true,
	linkcolor=red,
	citecolor=blue,
}
\newcommand{\red}[1]{\textcolor{red}{#1}}

\allowdisplaybreaks[2]

\usepackage{graphicx}
\usepackage{dcolumn}
\usepackage{bm}
\begin{document}

\preprint{APS/123-QED}

\title{Characterizing the effect of eccentricity on the dynamics of binary black hole mergers in numerical relativity}

\author{Hao Wang}
\email{husthaowang@hust.edu.cn}
\affiliation{Department of Astronomy, School of Physics, Huazhong University of Science and Technology, Wuhan 430074, China}

\author{Yuan-Chuan Zou}
\email{zouyc@hust.edu.cn}
\affiliation{Department of Astronomy, School of Physics, Huazhong University of Science and Technology, Wuhan 430074, China}

\author{Qing-Wen Wu}
\email{qwwu@hust.edu.cn}
\affiliation{Department of Astronomy, School of Physics, Huazhong University of Science and Technology, Wuhan 430074, China}

\author{Yu Liu}
\email{yuliu@gzu.edu.cn}
\affiliation{State Key Laboratory of Public Big Data, Guizhou University, Guiyang 550025, China}

\author{Xiaolin Liu}
\email{shallyn.liu@foxmail.com}
\affiliation{Department of Astronomy, Beijing Normal University, Beijing 100875, China}

\date{\today}

\begin{abstract}
Many articles have partially studied the configuration of eccentric orbital binary black hole (BBH) mergers. However, there is a scarcity of systematic and comprehensive research on the effect of eccentricity on BBH dynamics. Thanks to the rich and numerous numerical relativistic simulations of eccentric orbital BBH mergers from RIT catalog, this paper aims to investigate the impact of initial eccentricity $e_0$ on various dynamic quantities such as merger time $T_{\text{merger}}$, peak luminosity $L_{\text{peak}}$ of gravitational waves, recoil velocity $V_f$, mass $M_f$, and spin $\alpha_f$ of merger remnants. We cover configurations of no spin, spin alignment, and spin precession, as well as a broad parameter space of mass ratio ranging from 1/32 to 1 and initial eccentricity from 0 to 1.
For non-spinning BBH with an initial coordinate separation of $11.3M$ ($M$ is the total mass of BBH), we make the first discovery of a ubiquitous oscillation in the relationship between dynamic quantities $L_{\text{peak}}$, $V_f$, $M_f$, $\alpha_f$, and initial eccentricity $e_0$. Additionally, at $24.6M$, we observe the same oscillation phenomenon in the case of mass ratio $q=1$, but do not see it in other mass ratios, suggesting that this oscillation will be evident in numerical simulations with sufficiently dense initial eccentricity. By associating the integer numbers of the orbital cycle of $N_{\text{orbits}}$ with the peaks and valleys observed in the curves depicting the relationship between the dynamic quantities and the initial eccentricity, we reveal the significant oscillatory behavior attributed to orbital transitions. This discovery sheds light on the presence of additional orbital transitions in eccentric BBH mergers, extending beyond the widely recognized transition from inspiral to plunge. We perform an analysis to understand the different behaviors exhibited by the dynamic quantities and attribute them to variations in the calculation formulas. Furthermore, we demonstrate that finely adjusting the initial eccentricity can lead to the remnant black hole becoming a Schwarzschild black hole in the case of spin alignment. In a comprehensive analysis that surpasses previous studies by encompassing cases of no spin, spin alignment, and spin precession, we reveal consistent variations in the correlation between dynamic quantities and initial eccentricity, regardless of the presence of spin. This discovery underscores the universality of the impact of eccentricity on BBH dynamics and carries profound implications for astrophysical research.
\end{abstract}
\maketitle

\section{Introduction}
Since the groundbreaking detection of the gravitational wave event GW150914 in 2015 \cite{LIGOScientific:2016aoc}, gravitational wave astronomy has entered a transformative era. Over time, gravitational wave detection has evolved into a routine practice. Ground-based gravitational wave detectors, namely LIGO \cite{LIGOScientific:2014qfs}, Virgo \cite{VIRGO:2014yos}, and KAGRA \cite{KAGRA:2018plz} (collectively known as LVK), have successfully observed and recorded 93 gravitational wave events \cite{LIGOcollaboration}. These events encompass a variety of sources, including binary black holes (BBH), black hole-neutron star (BHNS) systems, and binary neutron star (NSNS) systems.

Since its breakthrough in solving the BBH merger problem \cite{Pretorius:2005gq,Campanelli:2005dd,Baker:2005vv}, numerical relativity (NR) has delved into deeper corners of the BBH parameter space. This technique has explored various scenarios, including systems with no spin, spin alignment, spin precession, eccentric orbits, and extreme mass ratios. However, most of the existing research in NR and gravitational wave detection has primarily focused on circular orbits. This emphasis on circularization is due to the gravitational wave radiation's circularizing effect \cite{Peters:1963ux,Peters:1964zz}, which eventually leads to BBH formed through the evolution of isolated binary stars in galaxy fields having circular orbits. These events of BBH mergers in circular orbits represent the primary targets for ground-based gravitational wave detectors such as LVK.
Nevertheless, there are several mechanisms through which BBH can acquire non-zero eccentricity before merging. In dense regions like globular clusters \cite{Miller:2002pg,Gultekin:2005fd,OLeary:2005vqo,Rodriguez:2015oxa,Samsing:2017xmd,Rodriguez:2017pec,Rodriguez:2018pss,Park:2017zgj} and galactic nuclei \cite{Gondan:2020svr,Antonini:2012ad,OLeary:2005vqo,Kocsis:2011jy,Hoang:2017fvh,Gondan:2017wzd,Samsing:2020tda,Tagawa:2020jnc}, BBH can gain eccentricity through processes \cite{OLeary:2007iqe} such as double-single interactions \cite{Samsing:2013kua,Samsing:2017oij}, double-double interactions \cite{Zevin:2018kzq,Arca-Sedda:2018qgq}, and gravitational capture \cite{Gondan:2020svr,East:2012xq}. Additionally, in three-body systems \cite{Naoz:2012bx} involving binary objects orbiting a supermassive black hole, the eccentricity of the inner binary can undergo oscillations due to the Kozai-Lidov mechanism \cite{VanLandingham:2016ccd,Silsbee:2016djf,Blaes:2002cs,Antognini:2013lpa,Naoz:2012bx,Stephan:2016kwj,Katz:2011hn,Seto:2013wwa}. These eccentric BBHs become detectable once they enter the frequency band of gravitational wave detectors. An example is the GW190521 event \cite{LIGOScientific:2020iuh}, which is considered a possible BBH merger with high mass and high eccentricity ($e=0.69_{-0.22}^{+0.17}$) \cite{Gayathri:2020coq,Romero-Shaw:2020thy}. With the continuous improvement in detector sensitivity, future ground-based gravitational wave detectors like the Einstein Telescope (ET) \cite{Punturo:2010zz} or the Cosmic Explorer (CE) \cite{Reitze:2019iox} are expected to detect an increasing number of eccentric BBH mergers.

Analytical relativity offers various methods to study the dynamics of BBHs merger, such as post-Newtonian (PN) \cite{Blanchet:2013haa}, effective one body (EOB) \cite{Buonanno:1998gg,Damour:2001tu}, and black hole perturbation theory (BHPT) \cite{Teukolsky:1973ha}. These analytical approaches are effective in describing the early adiabatic inspiral phase of BBHs. However, they fall short in capturing the extreme relativistic and nonlinear strong field dynamics, including the plunge and merger stages. To understand these crucial phases, we must rely on NR.
During the past decades, several NR collaborations, such as SXS \cite{Mroue:2013xna,Boyle:2019kee}, RIT \cite{Healy:2017psd,Healy:2019jyf,Healy:2020vre,Healy:2022wdn}, BAM \cite{Hamilton:2023qkv,Bruegmann:2006ulg,Husa:2007hp}, and MAYA \cite{Jani:2016wkt,Ferguson:2023vta}, have conducted numerous simulations of binary compact objects. They have made their simulation catalogs publicly available, contributing significantly to the field.

Modeling dynamical quantities such as peak luminosity, recoil velocity, remnant mass, and spin of BBH mergers carries significant astrophysical implications. However, due to the complexity of eccentric orbits, most articles that model these dynamical quantities mainly focus on circular orbits. In the case of precession, Ref. \cite{Taylor:2020bmj} employed the Gaussian process regression (GPR) method to model the peak luminosity.
Early estimations of recoil velocity relied on analytical approximation methods, including PN \cite{Blanchet:2005rj,Racine:2008kj}, EOB \cite{Damour:2006tr}, and closed limit approximation \cite{Sopuerta:2006wj}. Nowadays, more methods involve direct fitting of formulas with NR data \cite{Healy:2016lce,Healy:2014yta,Lousto:2009mf,Campanelli:2007ew,Lousto:2012gt,Hemberger:2013hsa,Healy:2018swt,Lousto:2007db}. Similarly, for the mass and spin of the remnant, fitting formulas with NR data \cite{Healy:2016lce,Healy:2014yta,Lousto:2009mf,Lousto:2013wta,Zlochower:2015wga,Lousto:2009ka,Lousto:2010xk,Keitel:2016krm,Healy:2017mvh}, analytical approximations \cite{Buonanno:2007sv}, and GPR \cite{Varma:2019csw,Varma:2018aht} are the commonly used methods.
Regarding NR simulations of eccentric orbits, there are currently limited open-source catalogs available, primarily including SXS \cite{SXSBBH}, MAYA \cite{Ferguson:2023vta} and RIT \cite{RITBBH}. The fourth release of RIT extends simulations to eccentric orbits, covering a wide parameter space \cite{Healy:2022wdn}. To date, only a few studies have explored the dynamic quantities in eccentric orbits, and most of them are qualitative in nature. These studies include investigating the influence of eccentricity on recoil velocity from a PN perspective \cite{Sopuerta:2006et}, analyzing the transition from inspiral to plunge in eccentric orbit \cite{Sperhake:2007gu}, studying orbital circularization \cite{Hinder:2007qu}, examining the recoil, mass and spin of remnant in low eccentricity orbits by NR \cite{Huerta:2019oxn}, exploring kick enhancement caused by eccentricity \cite{Sperhake:2019wwo}, and investigating anomalies in recoil due to eccentricity \cite{Radia:2021hjs}.
In an attempt to quantitatively model the remnant properties of low-eccentricity BBH mergers, Ref. \cite{Islam:2021mha} explores the use of GPR technology. Analytical modeling of these properties is challenging due to the added complexity introduced by eccentricity.

This paper aims to uncover the intricate nature of the complexity introduced by eccentricity, which may exceed our initial expectations. However, this complexity also opens the door to future analytical modeling.
RIT \cite{RITBBH} has conducted extensive and diverse simulations of eccentric orbit BBH mergers, which covered a wide range of parameters. These simulations include various mass ratios, ranging from 1/32 to 1, eccentricities spanning from 0 to 1, and consider scenarios with no spin, spin alignment, and spin precession. We provide a comprehensive summary of the relationships between the dynamic quantities of the merger time $T_{\text{merger}}$, peak luminosity $L_{\text{peak}}$, recoil velocity $V_f$, mass $M_f$, spin $\alpha_f$ of the merger remnants and the initial eccentricity $e_0$ in these three scenarios. Our study provides a systematic investigation and comprehensive analysis of the behavior exhibited by these quantities as the initial eccentricity varies.

This article is organized as follows. In Section \ref{sec:II}, we provide a summary of the numerical methods employed, the NR simulation data utilized for eccentric orbits, and introduce key concepts related to gravitational waves. In Section \ref{sec:III}, we present the NR data for two scenarios: the initial coordinate separation of $11.3M$ in Section \ref{sec:III:A:1} and the initial coordinate separation of $24.6M$ in Section \ref{sec:III:A:2}. Furthermore, we conduct an analysis of the observed behavior of the dynamic quantities in Section \ref{sec:III:A:3}. Then, we make a summary in Section \ref{sec:III:A:4}. In Section \ref{sec:III:B}, we explore the relationship between the dynamic quantities and the initial eccentricity for spin alignment, providing a detailed analysis and summary. Additionally, in Section \ref{sec:III:C}, we investigate spin precession case. Finally, in Section \ref{sec:IV}, we present our conclusions and provide an outlook for future research.
Throughout this article, we adopt geometric units where $G=c=1$. The component masses of BBH are represented as $m_1$ and $m_2$, while the total mass is denoted by $M$. For simplicity, we set the total mass $M$ at unity (although occasionally we explicitly write it for clarity). The mass ratio $q$ is defined as $q = m_1/m_2$, where $m_1$ is smaller than $m_2$. The dimensionless spin vectors of the black holes are denoted as $\vec{\chi}_i=\vec{S}_i / m_i^2$ for $i=1,2$, where $\vec{S}_i$ is spin vector of BBH.

\section{Eccentric Numerical Simulations}\label{sec:II}
The numerical relativistic simulations of eccentric orbital BBH mergers utilized in this study are obtained from the Rochester Institute of Technology (RIT) catalog. These simulations in the RIT catalog were performed using the LazEv code \cite{Zlochower:2005bj}, which implements the moving puncture approach \cite{Campanelli:2005dd} and employs the BSSNOK formalism for evolution systems \cite{Nakamura:1987zz,Shibata:1995we,Baumgarte:1998te} (except for cases involving highly spinning black holes, where the CCZ4 formalism \cite{Alic:2011gg} is used).
The LazEv code is integrated within the CACTUS/CARPET \cite{Schnetter:2003rb} infrastructure, which is part of the Einstein Toolkit \cite{Loffler:2011ay}. To locate apparent horizons, RIT employs AHFinderDirect \cite{Thornburg:2003sf}. Initially, RIT measures the amplitude of the horizon spins, denoted as $S_H$, utilizing the isolated horizon algorithm. Subsequently, they calculated the horizon mass using the Christodoulou formula: $m_H=\sqrt{m_{\mathrm{irr}}^2+S_H^2 /\left(4 m_{\mathrm{irr}}^2\right)}$, where $m_{\mathrm{irr}}$ represents the irreducible mass, defined as $m_{\mathrm{irr}}=\sqrt{A_H /(16 \pi)}$, with $A_H$ denoting the surface area of the horizon \cite{Campanelli:2006fy}.

For generating the numerical initial data, RIT employs the puncture approach \cite{Brandt:1997tf} in conjunction with the TwoPunctures code \cite{Ansorg:2004ds}. To determine the initial coordinate separation and tangential quasicircular momentum $p_{t,qc}$ for each eccentric family, RIT utilizes PN techniques as described in \cite{Healy:2017zqj}. By introducing a new parameter $\epsilon$, ranging from 0 to 1, the tangential linear momentum is modified as $p_t=p_{t,qc}(1-\epsilon)$. In this approach, the initial positions of BBH are fixed at the apocenter, and the initial orbital eccentricity gradually increases throughout the simulations, spanning from the quasi-circular orbit ($e=0$) to the head-on collision limit ($e=1$). The corresponding initial orbital frequency (and the (2,2)-modes of the gravitational waves) is reduced by the same factor $\Omega_e=\Omega_{qc}(1-\epsilon)$. Consequently, the initial eccentricity of the orbit can be approximated by $e=2\epsilon-\epsilon^2$, which provides a second order approximation in terms of $\epsilon$ and correctly captures the limits of $e=0$ and $e=1$ at $\epsilon=0$ and $\epsilon=1$, respectively.

RIT provides waveform data in the form of the Newman-Penrose scalar $\Psi_4$ and the gravitational wave strain $h$ which can be downloaded in RIT's catalog \cite{RITBBH}. These waveforms can be expanded using the spin-weighted spherical harmonic function ${ }_{-2}Y_{l,m}(\theta, \phi)$ with spin weight $s=-2$. Specifically, we have the expansion:
\begin{equation}\label{eq:1}
r\Psi_4=\sum_{l, m}r\Psi_4^{l m}{ }_{-2} Y_{l,m}(\theta, \phi),
\end{equation}
and
\begin{equation}
rh=r\left(h_{+}-i h_{\times}\right)=\sum_{l, m}rh_{l m}{ }_{-2} Y_{l, m}(\theta, \phi),
\end{equation}
where $r$ represents the extraction radius, $h_{+}$ and $h_{\times}$ denote the two polarizations of gravitational waves, and $h_{l m}$ and $\Psi_4^{lm}$ represent higher harmonic modes for $h$ and $\Psi_4$  respectively.
Furthermore, we can recall that the gravitational wave strain $h$ can be decomposed into a combination of amplitude and phase as follows:
\begin{equation}\label{eq:3}
h_{lm}=\mathcal{A}_{lm}(t) \exp \left[-i \Phi_{lm}(t)\right],
\end{equation}
where the amplitude $\mathcal{A}_{lm}$ and phase $\Phi_{lm}$ of $h_{lm}$ can be obtained using the following equations:
\begin{equation}\label{eq:4}
\mathcal{A}_{lm}=|h_{lm}|,
\end{equation}
\begin{equation}\label{eq:5}
\Phi_{lm}=arg(h_{lm}).
\end{equation}
To facilitate the representation of the parameter space and research, we introduce the concept of effective spin in the $z$ direction, which is aligned with the orbital angular momentum $L$. It is defined as
\begin{equation}
\chi_{\mathrm{eff}}=\frac{m_1 \chi_{1,z}+m_2 \chi_{2,z}}{m_1+m_2},
\end{equation}
where $\chi_{1,z}$ and $\chi_{2,z}$ represent the dimensionless spins of the two black holes in the $z$ direction. This measure allows us to characterize the combined spin of the binary system, considering the individual spins weighted by the respective masses of the black holes.
To quantify the precession effect, we adopt the effective precession spin parameter introduced in Ref. \cite{Schmidt:2014iyl}, defined as:
\begin{equation}\label{eq:7}
\chi_p=\frac{S_p}{A_1 m_1^2}.
\end{equation}
Here, we have the following:
\begin{equation}
\begin{aligned}
S_p & :=\frac{1}{2}\left(A_1 S_{1 \perp}+A_2 S_{2 \perp}+\left|A_1 S_{1 \perp}-A_2 S_{2 \perp}\right|\right) \\
& \equiv \max \left(A_1 S_{1 \perp}, A_2 S_{2 \perp}\right),
\end{aligned}
\end{equation}
where $S_{i \perp}$ ($i=1,2$) represents the component of the spin perpendicular to the orbital angular momentum. The values of $A_1$ and $A_2$ are given by $A_1=2+3 / 2q$ and $A_2=2+3q /(2)$, respectively.

The RIT catalog offers a comprehensive dataset that includes both waveform data and accompanying metadata, providing valuable information about the simulations. The metadata encompasses essential details regarding the initial data of the simulation, including mass ratio, initial distance, initial linear momentum, initial angular momentum, and more. Additionally, the metadata contains pertinent simulation results, such as the final remnant black hole masses, spins, and recoil velocity.
It is worth noting that these relaxed initial quantities are measured at a specific time, specifically $t_{\text{relax}}=200 M$, after the initial burst of radiation has substantially dissipated, accounting for relevant physical considerations.
To facilitate data exploration and visualization, RIT has organized all the information in an interactive table, ensuring convenient access and interpretation of the data set \cite{RITBBH}.

RIT employs formulas derived from Refs. \cite{Campanelli:1998jv,Lousto:2007mh} to quantify the radiated energy, linear momentum, and angular momentum using the radiative Weyl scalar $\Psi_4$. However, instead of utilizing the full $\Psi_4$, RIT decomposes it into $l$ and $m$ modes and focuses on the radiated linear momentum as in Eq. (\ref{eq:1}), disregarding terms with $l>6$. The resulting final recoil velocity is determined by linear momentum radiation.
In all simulations conducted by RIT, it has been ascertained that the waveforms, at the resolutions provided in the catalog, have reached a state of convergence, exhibiting convergence up to 4th-order with resolution. The evaluation of quantities related to the black hole horizon, such as the final mass and spins of the remnant, yields errors on the order of 0.1\% via the isolated horizon algorithm. Furthermore, radiatively computed quantities, including recoil velocities and peak luminosities, are evaluated with a typical error of 5\%.

\begin{figure*}[htbp!]
\centering
\includegraphics[width=15cm,height=7cm]{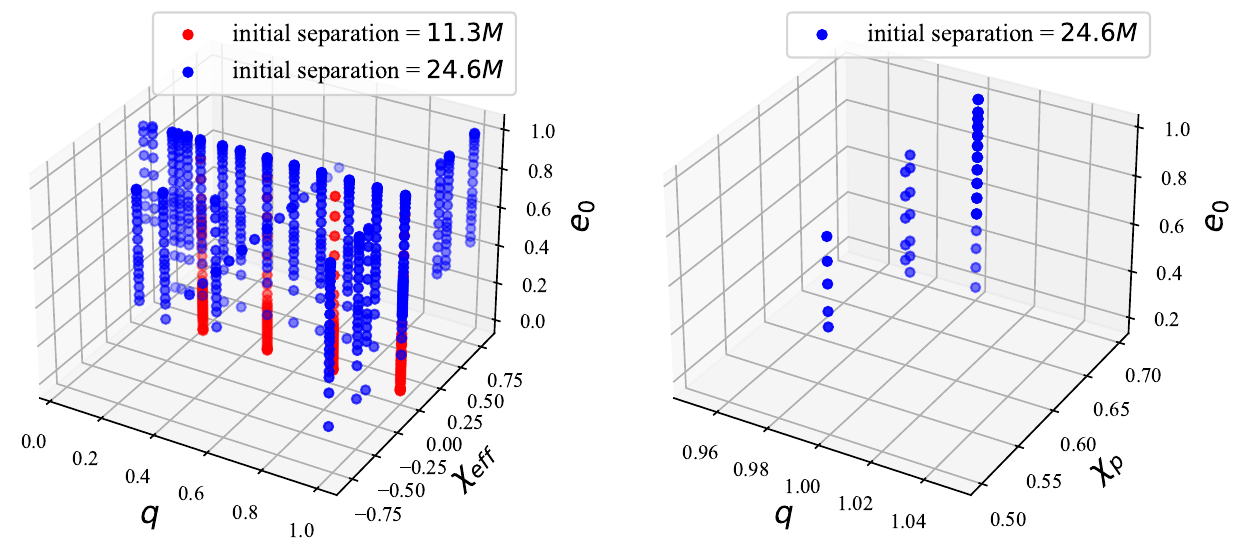}
\caption{\label{FIG:1}The parameters used in our study include three configurations no spin, spin alignment, and spin precession in two initial coordinate separations $11.3M$ and $24.6M$, which cover the parameter space mass ratio $q$ from 1/32 to 1, and the initial eccentricity $e_0$ from 0 to 1. The left panel uses effective spin $\chi_{\text{eff}}$ to label nonspinning and spin-aligned configurations. The right panel marks the spin precession configuration using the effective precession spin parameter $\chi_p$.}
\end {figure*}

The RIT catalog encompasses a broad range of numerical relativistic simulations, specifically focusing on eccentric orbital BBH mergers. The fourth release of the catalog introduces an extension to include eccentric orbits, featuring a total of 824 eccentric orbital BBH merger simulations. These simulations encompass a diverse parameter space, spanning eccentricities from 0 to 1, mass ratios ranging from 1/32 to 1, and various configurations, including nonspinning, spin-aligned, and spin-precessing setups.
It is worth noting that certain simulations were excluded from our research due to incomplete metadata, such as RIT:eBBH:1900 missing peak luminosity and recoil velocity, or the absence of a continuous sequence of eccentricity simulations like RIT:eBBH:1615 to RIT:eBBH:1620. However, these excluded data do not impact the results significantly. In the case of the former, the number of excluded simulations is minimal within the series of eccentric simulations, and they can be substituted with alternative simulations exhibiting similar eccentricities. As for the latter, it falls outside the scope of our study since the eccentricities in that particular series are not continuous. All simulations in our study is reported as both coordinate separaiton and proper distance. The initial coordinate separation, representing the coordinate separation between the two centroids of the black holes, is used to characterize the initial distance in our research. The two chosen initial coordinate separations are $11.3M$ and $24.6M$.
The parameter spaces for all simulations utilized in our study are depicted in FIG. \ref{FIG:1}. Specifically, we employed a total of 816 eccentric orbital BBH simulations, comprising 510 nonspinning, 197 spin-aligned, and 109 spin-precessing cases. The initial eccentricity, denoted as $e_0$, was estimated by RIT catalog and adopted as a reasonable approximation based on our earlier description.
For ease of visualization, FIG. \ref{FIG:1} presents the nonspinning, spin-aligned, and spin-precessing simulations in separate panels, employing the effective spin $\chi_{\mathrm{eff}}$ and effective spin precession parameters $\chi_p$ as characterization metrics.

\section{results}\label{sec:III}
Performing numerical relativity simulations is a computationally demanding task. Fortunately, the RIT catalog has undertaken a significant number of meticulous simulations focusing on eccentric BBH mergers. These simulations provide us with invaluable insights into the role of eccentricity in BBH merger dynamics. In this section, we present variations of various dynamical quantities, including the merger time $T_{\text{merger}}$, peak luminosity $L_{\text{peak}}$, recoil velocity $V_f$, mass $M_f$, and spin $\alpha_f$ of the merger remnants, as functions of the initial eccentricity $e_0$.
It is important to note that the merger time $T_{\text{merger}}$ represents the duration from $T_{\text{relax}}$ to the time of the peak of the gravitational wave amplitude. We analyze and discuss the behavior exhibited by these dynamical quantities, shedding light on their implications for eccentric BBH mergers.
However, it is regrettable that the trajectory information of the BBH system is not provided in the RIT catalog, which limits our ability to study other dynamic quantities, such as the evolution of the coordinate separation $D(t)$.

\subsection{No spin} \label{sec:III:A}
RIT catalog has conducted an extensive set of simulations focusing on no spin configurations, encompassing two different initial coordinate separations $11.3M$ and $24.6M$. Specifically, there are 191 simulation groups performed with an initial coordinate separation of $11.3 M$, and 319 groups with an initial coordinate separation of $24.6M$. The simulations with the former distance cover a finer range of initial eccentricities, while the simulations with the latter distance encompass a broader range of mass ratios. In this section, we will show the relationship between the dynamic quantities and the initial eccentricity in the nonspinning case and analyze them.

\subsubsection{ Initial coordinate separation = 11.3M}\label{sec:III:A:1}
\begin{figure*}[htbp!]
\centering
\includegraphics[width=15cm,height=15cm]{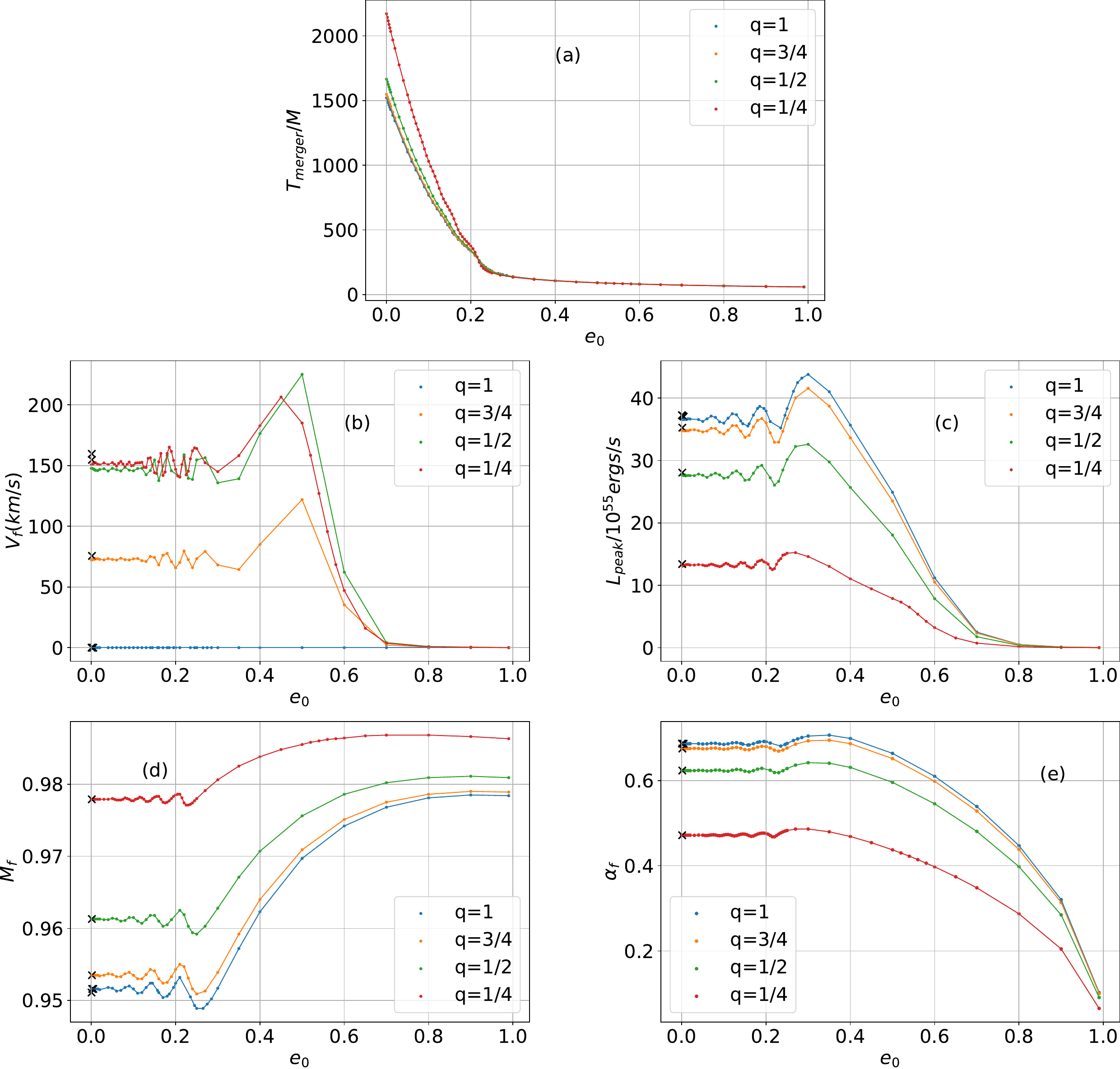}
\caption{\label{FIG:2}Variations of dynamical quantities of the merger time $T_{\text{merger}}$ (panel (a)), peak luminosity $L_{\text{peak}}$ (panel (c)), recoil velocity $V_f$ (panel (b)), mass $M_f$ (panel (d)), and spin $\alpha_f$ (panel (e)) of the merger remnants as a function of the initial eccentricity $e_0$ at the initial coordinate separation of $11.3M$ for nonspinning configuration with different mass ratio. Dynamic quantities of BBH mergers for various circular orbits are denoted by black marks ``x". These circular orbits correspond to RIT:BBH:0001 ($q=1$), RIT:BBH:0112 ($q=1$), RIT:BBH:0198 ($q=1$), RIT:BBH:0114 ($q=3/4$), RIT:BBH:0117 ($q=1/2$), and RIT:BBH:0119 ($q=1/4$) in the RIT catalog, respectively. Their initial orbital separations are 9.53, 20.0, 11.0, 11.0, 11.0, and 11.0, respectively.}
\end {figure*}

RIT catalog has conducted detailed simulations for the case where the initial coordinate separation is $11.3M$, focusing on specific mass ratios. In particular, RIT has performed fine simulations for the following mass ratios: 1/4 (67 groups), 1/2 (43 groups), 3/4 (41 groups), and 1 (41 groups). The emphasis on the number of simulation groups is of particular importance, as it significantly influences the presentation of the results.
In FIG. \ref{FIG:2}, we present the dynamical quantities of the merger time $T_{\text{merger}}$, peak luminosity $L_{\text{peak}}$, recoil velocity $V_f$, mass $M_f$, and spin $\alpha_f$ of the merger remnants, illustrating their variations as a function of the initial eccentricity $e_0$ at the initial coordinate separation of $11.3M$.

In panel (a) of FIG. \ref{FIG:2}, we present the evolution of the merger time $T_{\text{merger}}$ as a function of the initial eccentricity $e_0$. It is evident that $T_{\text{merger}}$ is influenced by two key factors: the mass ratio and the initial eccentricity. When the initial eccentricity $e_0$ is below approximately 0.23, $T_{\text{merger}}$ experiences a rapid decrease with increasing $e_0$. In contrast, when $e_0$ exceeds 0.23, $T_{\text{merger}}$ remains generally below $300M$ and gradually decreases, approaching zero.
It is important to note that the value $T_{\text{merger}}=0.23$ does not correspond to any specific dynamic positions, such as the transition from orbit to plunge. Furthermore, the influence of the mass ratio $q$ on $T_{\text{merger}}$ is evident, as smaller mass ratios result in longer $T_{\text{merger}}$ durations.

In panels (b), (c), (d), and (e) of FIG. \ref{FIG:2}, we illustrate the variations of the recoil velocity $V_f$, peak luminosity $L_{\text{peak}}$, mass $M_f$, and spin $\alpha_f$ as functions of the initial eccentricity $e_0$. Across all four panels, we observe a similar pattern in the behavior of these dynamic quantities.
Initially, these quantities are constant and are located in nearly horizontal straight lines. Then they display oscillatory behavior, which subsequently intensifies to a maximum or minimum value before eventually converging towards certain values in the head-on limit. Notably, the oscillations in peak luminosity $L_{\text{peak}}$, mass $M_f$, and spin $\alpha_f$ exhibit relatively regular and similar patterns. However, the oscillations in recoil velocity $V_f$ appear more chaotic and less predictable.
In a related study, Ref. \cite{Radia:2021hjs} identified a similar oscillation phenomenon in the recoil velocity during a series of numerical simulations for eccentric orbit BBH mergers at short initial separation. They referred to this phenomenon as ``anomalies" and attributed it to the infalling direction of the binary black holes at merger as a potential cause for this observed behavior.

In panel (b), the absence of linear momentum radiation in the case of $q=1$, which represents a completely symmetrical non-spinning configuration, results in a recoil velocity of 0. Ref. \cite{Gonzalez:2006md} discovered that for non-spinning circular orbits, the largest gravitational recoil occurs around $q=0.3$. Consequently, the recoil velocity for $q=1/4$ in panel (b) is higher compared to other mass ratios, because it is closest to 0.3.
When the initial eccentricity $e_0$ falls within the range of $[0,0.12]$, the oscillation of the recoil velocity is minimal, almost negligible. In the range of $[0.12,0.24]$, the recoil velocity exhibits a moderately chaotic oscillation. As eccentricity $e_0$ increases within the range of $[0.24,0.5]$, the recoil velocity experiences a sharp increase, reaching its maximum value. For $e_0$ in the range of $[0.5,0.99]$, the recoil velocity gradually decreases from the maximum value to 0 at the head-on collision limit. This characteristic holds true for all three mass ratios.

In panel (c), we observe a more regular oscillatory behavior compared to the recoil velocity, and it occurs earlier in the evolution. For the mass ratio $q=1$, there is almost no oscillation when the eccentricity $e_0$ ranges from $0$ to $0.05$. As the eccentricity increases within the range of $[0.05, 0.3]$, the oscillation gradually emerges and intensifies, reaching its maximum value. When the eccentricity is within the range of $[0.3, 0.99]$, the peak luminosity gradually decreases from the maximum value to the minimum value.
Furthermore, panel (c) reveals that the onset of the oscillation is delayed as the mass ratio decreases. The oscillation for the mass ratio $q=1/4$ begins at $e_0=0.095$, but $q=1$ starts earlier. And we can see as the mass ratio decreases, the oscillation becomes weaker. Additionally, smaller mass ratios correspond to lower peak luminosities, consistent with the behavior observed in circular orbits. Another noteworthy observation is that the initial eccentricity corresponding to the maximum value of the oscillation increases as the mass ratio increases.

In panel (d), we observe a similar oscillatory behavior in the mass of the remnant with respect to the eccentricity $e_0$, resembling the pattern seen in the peak luminosity. However, the opening corresponding to the peak is oriented upwards. For the mass ratio $q=1$, there is almost no oscillation when the eccentricity $e_0$ is in the range of $[0, 0.02]$. As the eccentricity increases within the range of $[0.02, 0.26]$, the oscillation emerges and gradually decreases to its minimum value. Notably, the onset of oscillation occurs earlier than that of the peak luminosity. When the eccentricity is within the range of $[0.26, 0.99]$, the mass of the remnant gradually increases from its minimum value to 0.98. It is important to mention that the mass of the remnant is not exactly \red{one} when the eccentricity $e_0=1$, as the binary black holes also radiate energy during the head-on collision, although the amount is relatively small.
Similar to the peak luminosity, the oscillatory behavior of the mass of the remnant becomes weaker as the mass ratio decreases. Furthermore, lower mass ratios correspond to smaller eccentricities at which the oscillation reaches its minimum value. On the contrary, larger mass ratios result in more energy being radiated, leading to a smaller mass of the remnant, which is consistent with the behavior observed in circular orbits.

In panel (e), we observe that the oscillation of the spin of the remnant $\alpha_f$, is significantly weaker compared to the recoil velocity $V_f$, peak luminosity $L_{\text{peak}}$, and mass $M_f$. For the mass ratio $q=1$, there is almost no oscillation when the eccentricity $e_0$ falls within the range of $[0, 0.05]$. As the eccentricity increases within the range of $[0.05, 0.35]$, the oscillation gradually emerges and intensifies, reaching its maximum value. Subsequently, when the eccentricity is in the range of $[0.35, 0.99]$, $\alpha_f$ gradually decreases from the maximum value to 0.1. This residual spin of 0.1 is likely a result of some remaining orbital angular momentum in the initial data.
The characteristics of the spin oscillation of the remnant with respect to the mass ratio exhibit similarities with those of the peak luminosity $L_{\text{peak}}$ and mass $M_f$, and will not be discussed here.

To ensure the comprehensiveness of our study, we present two new approaches for estimating eccentricity in Appendix \ref{App:B}. This method utilizes ADM coordinates and harmonic coordinates, derived from the generalized quasi-Keplerian parameterization of 3PN \cite{Memmesheimer:2004cv}. Furthermore, we conduct a comparative analysis of three distinct eccentricity measurement methods (ADM coordinates, harmonious coordinates, and RIT approximation methods).

\subsubsection{Initial coordinate separation = 24.6M} \label{sec:III:A:2}
\begin{figure*}[htbp!]
\centering
\includegraphics[width=15cm,height=15cm]{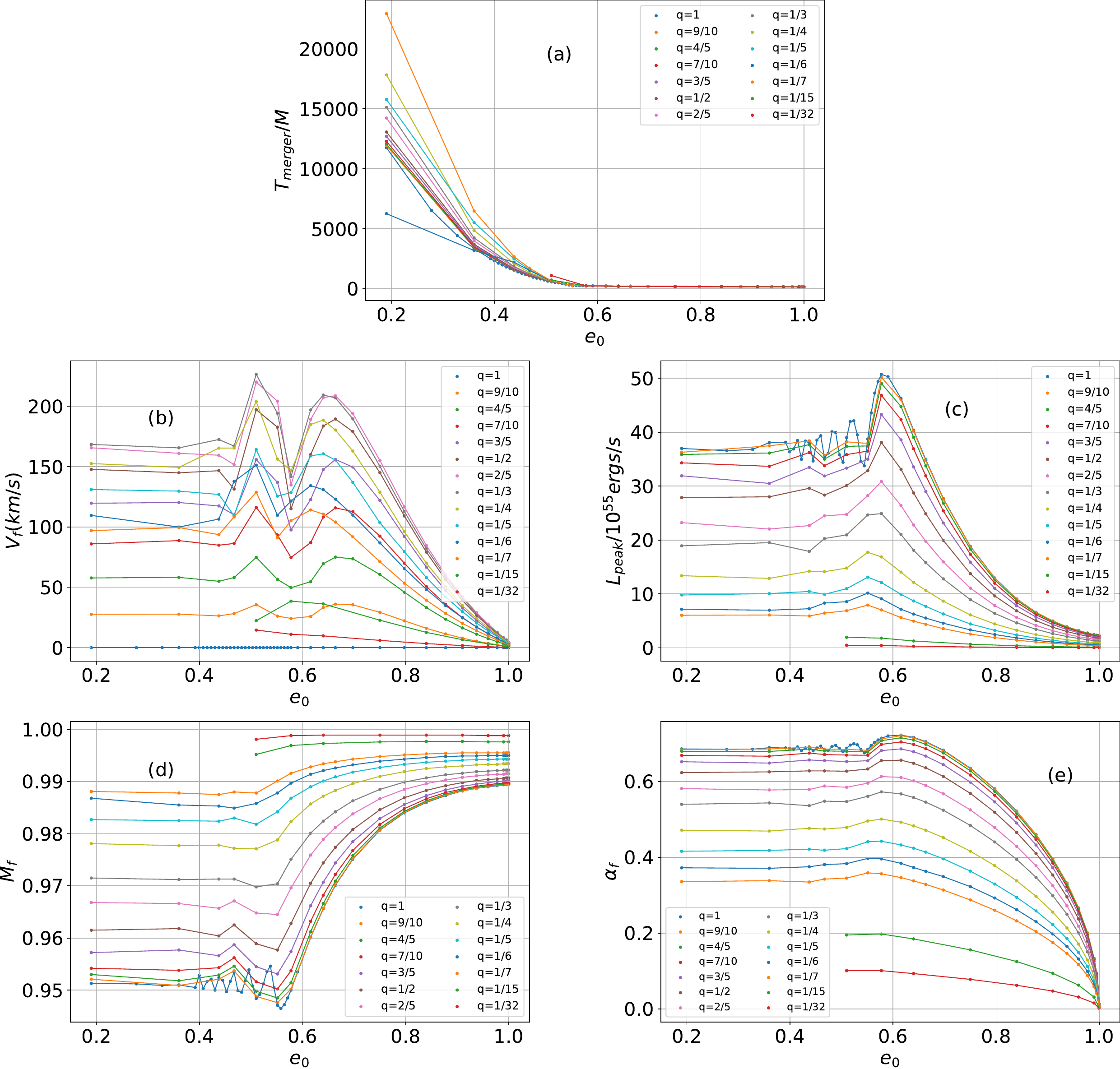}
\caption{\label{FIG:3}Variations of dynamical quantities of the merger time $T_{\text{merger}}$ (panel (a)), peak luminosity $L_{\text{peak}}$ (panel (c)), recoil velocity $V_f$ (panel (b)), mass $M_f$ (panel (d)), and spin $\alpha_f$ (panel (e)) of the merger remnants as a function of the initial eccentricity $e_0$ at the initial coordinate separation of $24.6M$ for nonspinning configuration with different mass ratio.}
\end {figure*}

Next, we examine a significantly larger initial coordinate separation of $24.6M$. RIT has conducted extensive simulations for various mass ratios for the case. The first is $q=1$ for 48 groups, which has the largest number of groups among them. Additionally, simulations were performed for mass ratios of 9/10, 4/5, 7/10, 3/5, 1/2, 2/5, 1/3, 1/4, 1/5, 1/6, and 1/7 for 23 groups, as well as 1/15 and 1/32 for 9 groups. Simulations of the last two mass ratios lack low and moderate eccentricities.
In FIG. \ref{FIG:3}, we present the variations of dynamical quantities of merger time $T_{\text{merger}}$, peak luminosity $L_{\text{peak}}$, recoil velocity $V_f$, mass $M_f$, and spin $\alpha_f$ of the merger remnants as a function of the initial eccentricity $e_0$ at the initial coordinate separation of $24.6M$.

In panel (a) of FIG. \ref{FIG:3}, we present the variation of the merger time length $T_{\text{merger}}$ as a function of the initial eccentricity $e_0$ for an initial coordinate separation of $24.6M$. In particular, the relationship between the merger time and the initial eccentricity exhibits a pattern similar to that observed in the case of $11.3M$.
Interestingly, we observe a distinct turning point at approximately $e_0=0.5$ for the $24.6M$ case, in contrast to the turning point at $e_0=0.23$ observed in the $11.3M$ case. This discrepancy can be attributed to the larger initial coordinate separation utilized in the $24.6M$ scenario. We believe this is a general behavior in which a larger initial coordinate separation leads to a delayed turning point.
Furthermore, our findings indicate that the merging time increases with decreasing mass ratio. This trend also holds for the $24.6M$ case as well. However, it is important to note that the mass ratio $q=0.1667$ shows a significant deviation from the other results due to errors present in the data itself. In our previous study \cite{Wang:2023ueg}, we discovered that the waveforms associated with a mass ratio of $q=1/6$ exhibit abnormal behavior and peculiar deviations from the expected patterns. The observed performance reveals a noteworthy discrepancy in the fitting parameters when comparing $q=1/6$ to other mass ratios. Additionally, simulation RIT:eBBH:1537 with $q=1/6$ exhibits an apparent issue in its waveform, possibly attributable to the center of mass drifting.

In panels (b), (c), (d), and (e) of FIG. \ref{FIG:3}, we present recoil velocity $V_f$, peak luminosity $L_{\text{peak}}$, mass $M_f$, and spin $\alpha_f$, as a function of the initial eccentricity $e_0$. Notably, the overall pattern observed in the curves aligns closely with the trends observed in the $11.3M$ case. However, there are notable differences, particularly in the presence or absence of oscillatory behavior among different groups of mass ratios.
In panel (b), an intriguing bimodal structure emerges in the relationship between recoil velocity and initial eccentricity. This unexpected pattern adds a fascinating layer to our understanding of the merger process and warrants further investigation.
It is worth highlighting that, while general trends remain consistent with the case $11.3M$, the oscillation behavior seen in the groups other than the mass ratio $q=1$ is less pronounced or absent. This discrepancy adds an intriguing dimension to the dynamics of the merger remnants and prompts us to explore the underlying mechanisms responsible for these observations.

In panel (b) of FIG. \ref{FIG:3}, we observe that the largest group with a mass ratio of $q=1$ exhibits complete symmetry, resulting in no linear momentum radiation and, consequently, a recoil velocity of 0. However, for other mass ratios, a visible oscillatory pattern emerges, commencing at an approximate eccentricity of 0.44. The first peak in the recoil velocity occurs at an initial eccentricity of 0.51. Notably, the position of the second peak is not fixed and varies with the mass ratio. Specifically, in the case of a mass ratio of $q=1/3$, the second peak manifests at an initial eccentricity of 0.64. Subsequently, the recoil velocity progressively decreases to 0 as the eccentricity increases.
Analyzing the results for the initial coordinate separation of $24.6M$, we find that, apart from the peculiar mid-range peaks and the subtle oscillation behavior, the overall trends align with those observed in the case of $11.3M$. Furthermore, the maximum recoil velocity occurs in a mass ratio of $q=1/3$, consistent with the findings for the scenario $11.3M$. Additionally, we reaffirm the pattern that smaller mass ratios correspond to smaller oscillation or peak values, further validating the observations from the $11.3M$ case and illustrating this trend across a wider range of mass ratios.
These findings not only deepen our understanding of recoil dynamics in merger remnants for an initial coordinate separation of $24.6M$, but also reinforce and extend the variations observed in the $11.3M$ case, providing valuable insights across a broader range of mass ratios.

In panel (c) of FIG. \ref{FIG:3}, we observe that the overall behavior of the peak luminosity $L_{\text{peak}}$ is consistent with the findings for the scenario $11.3M$. Initially, it remains relatively constant, followed by oscillations that gradually reach a maximum value. Subsequently, the luminosity decreases from its peak to a minimum value.
Notably, we clearly observe the oscillation pattern in the case of a mass ratio of $q=1$, which is supported by a robust dataset of 48 groups. However, the oscillation behavior is less apparent for other mass ratios, where the available data is only half the size, comprising 23 groups. The sparser data points for mass ratios other than $q=1$ smooth out the oscillations due to the coarse-graining of the initial eccentricity, making them less discernible.
Furthermore, we find that for the mass ratio $q=1$, the oscillations begin at a higher initial eccentricity of 0.33. In contrast, we observe only subtle undulations and peaks for other mass ratios. Despite the absence of clear oscillations for mass ratios other than $q=1$, we can draw analogous conclusions based on a wider range of mass ratios, similar to the $11.3M$ case. Specifically, we find that smaller mass ratios correspond to lower peak luminosity, weaker oscillations, and a shift in the position of the maximum oscillation towards lower initial eccentricities.
It is important to note that these conclusions are drawn by analogy since we do not observe explicit oscillations for mass ratios other than $q=1$. However, based on the findings in the $11.3M$ case, we would expect such oscillations to exist given a sufficient number of data points. To achieve deterministic oscillations resembling those observed in the $11.3M$ case, further numerical relativistic simulations of eccentric orbits with an initial separation of $24.6M$ would be required. However, it is important to note that such simulations would entail a substantial increase in computational cost.

In panel (d) of FIG. \ref{FIG:3}, we present the variation of the remnant mass with respect to the initial eccentricity. The overall behavior closely resembles that of the peak luminosity, exhibiting similar trends. However, for the sake of brevity, we will refrain from delving into further details in this section.

In panel (e) of FIG. \ref{FIG:3}, we observe that the oscillation of the remnant spin is comparatively weaker than the oscillations observed in the other three dynamic quantities, aligning with the findings of the $11.3M$ scenario.
Additionally, we note that as the mass ratio decreases, the maximum value of the oscillation shifts to lower initial eccentricities, consistent with the observations in the $11.3M$ case. The analysis of other relationships exhibits similar patterns, and we refrain from reiterating them here to avoid redundancy.

In summary, our study involves a series of numerical simulations in which we systematically increased the initial eccentricity, while maintaining a fixed initial coordinate separation of $11.3M$ or $24.6M$. From these simulations, we derive several dynamic quantities characterizing the merger process. We observe consistent behaviors of the merger time $T_{\text{merger}}$, peak luminosity $L_{\text{peak}}$, recoil velocity $V_f$, mass $M_f$, and spin $\alpha_f$ of the merger remnants with changes in the initial eccentricity for both cases.
The merger time $T_{\text{merger}}$ exhibits an initial rapid decrease, followed by a slower decrease after passing a critical point. On the other hand, the remaining four quantities, peak luminosity $L_{\text{peak}}$, recoil velocity $V_f$, mass $M_f$, and spin $\alpha_f$ of the merger remnants, display a universal behavior with the changing initial eccentricity. Initially, they maintain an almost stable horizontal line. Subsequently, they gradually enter an oscillatory phase, with the amplitude of oscillations intensifying. At the final peak, these quantities reach a maximum or minimum value. Finally, under extreme eccentricity $e_0=1$ or in the head-on collision limit, they gradually approach a specific value. This behavior is only observable when the initial eccentricity in the numerical simulation is sufficiently dense.
Among these dynamic quantities, the oscillation behavior of the recoil velocity $V_f$ appears relatively irregular and less ordered compared to the relatively regular oscillations observed in $L_{\text{peak}}$, $M_f$, and spin $\alpha_f$. Furthermore, the magnitude of the oscillations decreases as the mass ratio decreases and the initial eccentricity corresponding to the maximum or minimum value shifts with changes in the mass ratio.

\subsubsection{Analysis}\label{sec:III:A:3}
Understanding the intricate relationship between merger time and initial eccentricity, as well as the impact of the initial coordinate distance and mass ratio, can be accomplished through the application of analytical PN theory \cite{Blanchet:2013haa}. Our investigation reveals a notable turning point in this relationship. When the eccentricity exceeds the critical value, the merger time tends to decrease, although at a slower pace. Visually, a gradual decline in merger time with increasing eccentricity is observed. For a comprehensive analysis, we refer the interested reader to the full publication of PN.

The primary focus of our investigation is to reveal the underlying physical mechanisms responsible for the observed oscillatory behavior in the dynamic quantities of peak luminosity $L_{\text{peak}}$, recoil velocity $V_f$, mass $M_f$, and spin $\alpha_f$ of merger remnants. Additionally, we aim to quantify the extent to which these quantities can be enhanced or diminished by manipulating the initial eccentricity. Through a detailed analysis, we endeavor to elucidate the fundamental factors driving these oscillations and provide insights into the potential impact of varying the initial eccentricity on these dynamic quantities.

In the study conducted by Huerta et al. \cite{Huerta:2019oxn}, a comprehensive set of 89 eccentric numerical relativistic simulations was performed. These simulations covered a wide range of mass ratios from 1 to 10, with corresponding eccentricities ranging from 0 to 0.18. Their analysis focused on establishing the relationship between the mass, spin, and recoil of the merger remnant and the initial eccentricity.
Interestingly, Huerta et al. found that these dynamic quantities exhibited minimal changes with respect to the initial eccentricity. Moreover, the dynamic quantities remained relatively constant, forming a nearly horizontal line in the parameter space. It is worth noting that the simulations conducted by Huerta et al. had initial coordinate separations exceeding $11.3 M$, as indicated by the number of cycles of gravitational waves. And they did not simulate enough eccentricity data points. Consequently, the specific range of eccentricities explored (up to 0.18) did not induce oscillatory behavior in the system.

Radia et al. \cite{Radia:2021hjs} conducted nonspinning eccentric numerical simulations with mass ratios of $q=1/2$, $1/3$, and $2/3$. Their research focused on investigating the recoil velocity of the merger remnant, where they observed intriguing oscillatory behavior. Additionally, their figures exhibited noticeable oscillations in the remnant's spin and radiated energy, although these quantities were not the primary focus of their study.
Of particular interest is their examination of cases involving very short initial coordinate separations, close to the point of merger. It is worth noting that their approach to generating eccentricity differs from that of RIT. Initially, they established a quasicircular configuration by fixing the binding energy $E_{\mathrm{b}}$, defined as
\begin{equation}\label{eq:9}
E_{\mathrm{b}}=M_{\mathrm{ADM}}-M,
\end{equation}
where $M_{\mathrm{ADM}}$ is ADM mass. Subsequently, they incrementally reduced the initial linear momentum parameter $p$ to generate a series of eccentric simulations. Importantly, while the eccentricity varied, the initial coordinate separation gradually increased. This finding offers an alternative perspective, demonstrating that the observed oscillation phenomenon is universal and independent of the initial distance in the simulation. 
This oscillatory phenomenon in the recoil velocity was explained by Radia et al. \cite{Radia:2021hjs} as a consequence of the change in infall direction during the BBH merger. However, it should be noted that this oscillation is not limited to the recoil velocity alone. The oscillatory behavior was also observed in the dynamic quantities of peak luminosity $L_{\text{peak}}$, recoil velocity $V_f$, mass $M_f$, and spin $\alpha_f$. Notably, the peaks and valleys of these oscillations for the dynamic quantities were situated at different initial eccentricity positions and did not correspond to each other. Furthermore, the oscillations exhibited both maximum and minimum values, suggesting a more complex underlying cause. These phenomena cannot be solely attributed to changes in the infall direction. The observed oscillations between the infall direction and the recoil are better characterized as a phenomenological outcome rather than as a representation of a singular physical origin.

Sopuerta et al. \cite{Sopuerta:2006et} provided a PN perspective, indicating that in the low eccentricity regime, the recoil velocity $V_f$ scales as $\propto\left(1+e_0\right)$. This PN analysis establishes a relationship between the recoil velocity and eccentricity within this specific regime.
Furthermore, references such as Radia et al. \cite{Radia:2021hjs} and Sperhake et al. \cite{Sperhake:2019wwo} demonstrate that nonzero eccentricity can lead to a significant increase in the recoil velocity $V_f$, up to approximately 25\% when compared to the quasi-circular orbit case. These studies provide valuable insights into the enhancement of recoil velocity resulting from the presence of eccentricity.
While there exist references that have investigated the enhancement of recoil velocity $V_f$ caused by eccentricity, such as those mentioned earlier, there is limited literature that delves deeply into the amplification of peak luminosity $L_{\text{peak}}$, mass $M_f$, and spin $\alpha_f$ induced by nonzero initial eccentricity. It is of utmost importance to acknowledge that eccentricity introduces a distinctive oscillatory behavior, resulting in both amplifications and reductions in these dynamic quantities, rather than exclusively leading to amplifications. Therefore, further exploration is necessary to gain a comprehensive understanding of the effects of nonzero eccentricity on peak luminosity, mass, and spin of the merger remnant.

Hinder et al. \cite{Hinder:2007qu} conducted a series of numerical simulations focusing on cases of high eccentricity. They analyzed the changes in spin and mass of the merger remnants and concluded that the orbit becomes circularized when the eccentricity drops below 0.4. It is important to note that the initial separations in their simulations were approximately $12M$, which is very close to the value of $11.3M$ used in this paper.
From the perspective of more refined eccentric numerical simulations conducted by RIT (refer to FIG. \ref{FIG:2}) and in light of the conclusions of Ref. \cite{Hinder:2007qu}, it is observed that the eccentricity reaches a lower value (specifically $e_0=0.02$) when the orbit starts to circularize completely. However, above this value, the orbit cannot be fully circularized. This conclusion we get may seem counterintuitive, as we typically associate circularization with some small eccentricities, but not as small as 0.02. However, it is crucial to distinguish the conceptual difference between the mass and spin of the remnant which are integral quantities, and the instantaneous circularization state of the orbit. While these two concepts can provide some characterization of each other, they are not entirely equivalent.
In fact, the process of circularization is also reflected in the oscillatory phenomena observed in the peak luminosity, recoil velocity, mass, and spin of the merger remnant. Weaker oscillations indicate a stronger degree of circularization. Notably, there is minimal oscillatory effect for initial eccentricities ranging from 0 to 0.02. This does not imply complete circularization at eccentricities below 0.02, but rather suggests that for $e_0\le0.02$, the dynamics and waveforms of these simulations closely resemble those of quasicircular orbits.

The appearance of oscillations in peak luminosity, recoil velocities, masses, and spins of merger remnants is an intriguing phenomenon. Understanding the origin of these oscillations is closely tied to the peaks observed within the oscillatory behavior, particularly the last peak, which tends to be the largest and introduces the most significant enhancement effect caused by eccentricity.
In a relevant study, Sperhake et al. \cite{Sperhake:2007gu} investigated the transition from inspiral to plunge in eccentric BBH mergers. They explored a wide range of eccentricities from 0 to 1 and examined the relationship between eccentricity and radiated energy. In particular, they found that near the critical point that marks the transition from orbit to plunge, the spin parameter $\alpha_f$ of the remnant reached a maximum value of 0.724. While their study provided valuable insights into the eccentricity dependence of the remnant's spin and its relation to the transition from inspiral to plunge, the oscillatory phenomenon was not observed due to the relatively small number of numerical simulations conducted, amounting to only a dozen sets.
Nevertheless, the findings presented in Ref. \cite{Sperhake:2007gu} offer valuable guidance for analyzing the underlying mechanisms responsible for the generation of oscillations in dynamic quantities.

In this study, we draw inspiration from the concept presented in Ref. \cite{Sperhake:2007gu} to consider the orbital transition. The number of orbital cycles $N$ can be determined from two perspectives: one through the orbital phase of the puncture and the other through the phase of the gravitational waveform. However, it should be noted that RIT does not provide orbit trajectory information, restricting our ability to calculate the number of orbital cycles $N$ solely through the gravitational waveform. In our analysis, we specifically focus on the 2-2 mode.
To calculate the phase difference, we evaluate the expression:
\begin{equation}\label{eq:10}
\Delta \Phi=\Phi\left(t_{\mathrm{merger}}\right)-\Phi\left(t_0+t_{\mathrm{relax}}\right).
\end{equation}
where $t_{\mathrm{merger}}$ represents the time of BBH merger, $t_0$ denotes the initial moment of the waveform, and $t_{\mathrm{relax}}$ signifies the time required to the transition from the initial moment to a physically stable state. For the phase calculation, we adopt $t_{\mathrm{relax}}=20M$ to remove small steps in the phase. While $t_{\mathrm{merger}}$ is determined as the time when a common apparent horizon is formed as used in Ref. \cite{Sperhake:2007gu}, we only have the time $t_{\mathrm{merger}}$ that corresponds to the maximum amplitude in the waveform data. However, employing $t_{\mathrm{merger}}$ instead of the precise time of common apparent horizon formation does not introduce a significant error in the phase difference calculation.
Subsequently, the number of orbital cycles accomplished by the BBH system can be obtained as:
\begin{equation}\label{eq:11}
N_{\text {orbits }}=\frac{\Delta \Phi}{4 \pi}.
\end{equation}
Here, we divide the phase difference $\Delta \Phi$ by $4 \pi$ since the waveform phase is twice that of the orbital phase. This conversion factor is chosen to align the two quantities appropriately.

\begin{figure*}[htbp!]
\centering
\includegraphics[width=15cm,height=20cm]{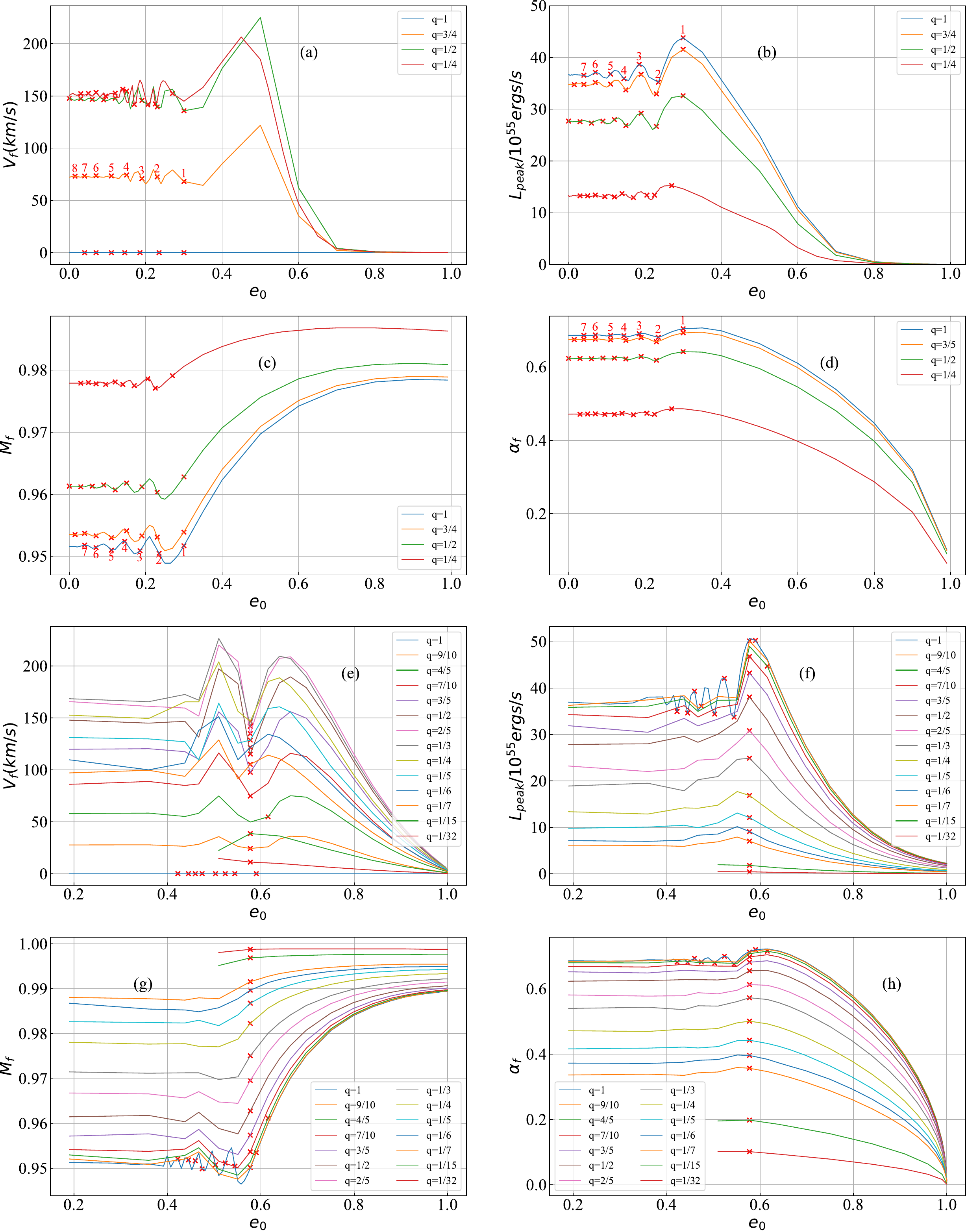}
\caption{\label{FIG:4}Relationship between the integer orbital cycle number $N_{\text{orbits}}$ and various quantities such as the peak luminosity $L_{\text{peak}}$, recoil velocity $V_f$, the mass $M_f$ and the spin $\alpha_f$ of the remnant at initial coordinate separation of $11.3M$ and $24.6M$ for nonspinning configuration with different mass ratios. These points, denoted by red ``x" markers, correspond to either an integer or are in close proximity to an integer of the orbital cycle number. Moving from right to left, each red ``x" corresponds to successive orbital cycles, starting from cycle 1 and continuing indefinitely. In the case of a mass ratio of $11.3M$ being $q=1$, we illustrate the integer orbital cycle numbers in the figure using numerical values 1, 2, 3 ... (with the exception of the recoil velocity, which is represented by $q=1/4$). The upper four panels correspond to the initial coordinate separation of $11.3M$, and the lower four panels correspond to the initial coordinate separation of $24.6M$.}
\end {figure*}

FIG. \ref{FIG:4} displays the relationship between the integer orbital cycle number $N_{\text{orbits}}$ and various quantities such as peak luminosity $L_{\text{peak}}$, recoil velocity $V_f$, mass $M_f$, and spin $\alpha_f$ of the remnant at the initial coordinate separation of $11.3M$ and $24.6M$ for different mass ratios. These points, denoted by red ``x" markers, correspond to either an integer multiple or are in close proximity to an integer multiple of the orbital cycles. In the case of a mass ratio of $11.3M$ being mass ratio $q=1$, we illustrate the integer orbital cycle numbers in the FIG. \ref{FIG:4} using numerical values 1, 2, 3 ... (with the exception of the recoil velocity, which is represented by $q=1/4$. For the case with an initial coordinate separation of $24.6M$, we accurately mark the points only for a mass ratio of $q=1$). Due to the limited number of numerical simulation groups available for other mass ratios at the initial coordinate separation of $24.6M$, there exist significant deviations from integer cycles or instances of excessive discontinuous cycles. Therefore, we selectively mark the cases where $N_{\text{orbits}}$ closely approximates 1 in FIG. \ref{FIG:4} for $24.6M$ with other mass ratios. As discussed earlier, the orbital cycle value obtained from the gravitational wave phase is not precisely an integer but may deviate to some extent. This deviation can lead to a mismatch between the integer orbital cycle and the observed peaks and valleys of the oscillation.  In order to ensure a comprehensive analysis, we present the error of the nearest continuous integer orbital cycle $N_{\text{orbits}}$ compared to the corresponding integer value for the $11.3M$ case in FIG. \ref{FIG:9} of Appendix \ref{App:A}. The maximum relative error is approximately 0.15, with the majority of errors concentrated below 0.1.

We now shift our focus to panels (b), (d), (f) and (h) in FIG. \ref{FIG:4}, which corresponds to the peak luminosity and spin of the remnant. These panels exhibit similarities to the scenarios investigated in Ref. \cite{Sperhake:2007gu}. In particular, we observe that all cases with $N_{\text{orbits}} = 1$ align precisely with the last peak, indicative of the transition from inspiral to plunge. Additionally, instances with $N_{\text{orbits}} = 2$ are predominantly positioned near the last valley. Furthermore, cases with $N_{\text{orbits}} = 3$ are consistently found in the penultimate peak, and this pattern continues for higher values of $N_{\text{orbits}}$.
We contend that this observed behavior is not coincidental but rather stems from a shared physical origin underlying the generation of both the last peak and other peaks or valleys. Much like how the last peak signifies the transition from inspiral to plunge, the last valley represents the transition from the last orbit 2 to the last orbit 1, while the penultimate peak corresponds to the transition from the last orbit 3 to the last orbit 2, and so on. Consequently, we gain insight into why dynamic quantities such as recoil velocity, peak luminosity, mass, and spin progressively oscillate from an initial horizontal line, culminating in a maximum peak or deepest valley. Moreover, we ascertain that the transition from inspiral to plunge introduces the most substantial enhancement effect in eccentric BBH mergers, aligning with the conclusion drawn in Ref. \cite{Sperhake:2007gu}.
This behavior holds for both an initial separation of $11.3M$ or $24.6M$ and mass ratios up to $q=1$. Nevertheless, in panels (b), (d), (f), and (h), certain data points deviate from the peaks or valleys, with larger deviations occurring the farther they are from the last few peaks. Several factors may contribute to these deviations:

(i) Due to the limited simulation data, we are unable to obtain an exact integer value for the cycle $N_{\text{orbits}}$, resulting in deviations from integers. The last few peaks and valleys are more apparent due to the fine simulation eccentricity, allowing for more accurate results. Occasionally, the worst cycle $N_{\text{orbits}}$ deviates from an integer by up to 0.15, leading to significant errors.

(ii) The data obtained from the simulations are not finely resolved but rather coarse-grained. Consequently, the peaks and valleys we identify may not precisely align with their most accurate positions but exhibit some level of deviation.

(iii) As mentioned in Sec. \ref{sec:II}, the peak luminosities, recoil velocities, masses, and spins that we obtain are subject to errors. Simultaneously, errors arise in the phase used to calculate the cycle number  $N_{\text{orbits}}$.

(iv) In eccentric BBH mergers, strong periastron precession occurs, causing the orbital plane to process similarly to the perihelion precession of Mercury \cite{Will:2014kxa}. This precession leads to an incomplete orbital phase of the BBH, deviating from $2\pi$. The greater the number of orbits and the smaller the mass ratio, the more severe the deviation. This effect may be a significant contributor to the observed discrepancies, where many data points do not exactly correspond to the peaks and valleys.

It is important to note that while the measurement of eccentricity may be subject to significant errors due to approximate measurement methods, these errors do not impact the position of the peak or valleys and the occurrence of oscillatory behavior, since the way in which the initial eccentricity being generated is continuous and physically reasonable. The presence of uncertainties in the eccentricity measurements does not alter the overall pattern observed in the data.
Furthermore, it is worth mentioning that there is no analytical formula available for the peak luminosity. However, the similarity observed between the peak luminosity and the spin of the remnant can be attributed to their inherent correlation, as discussed in Ref. \cite{Ferguson:2019slp}.

Moving on to the mass of the remnant $M_f$, it is evident that many integer orbital cycles do not align precisely with peaks or valleys. Rather, there are some specific deviations. This behavior can be likened to a phase shift, where the remnant mass is shifted in phase relative to the initial eccentricity. This phase shift arises due to the specific calculation method employed to determine $M_f$, which differs somewhat from the calculations for peak luminosity and spin. Although the integer cycle points for $M_f$ do not coincide with the peaks and valleys, we observe that the differences in the cycle number $N_{\text{orbits}}$ between the peaks or valleys of the remnant mass $M_f$ are approximately 1 when calculated. This finding highlights that the emergence of peaks and valleys in the remnant mass is a result of transitions between orbits, mirroring the behavior observed in peak luminosities and spins. This also shows that our conclusion that the oscillations come from an orbital transition is self-consistent.

Before we proceed with the analysis of recoil velocity, let us first recall some formulas from Refs. \cite{Ruiz:2007yx,Radia:2021hjs} that are used to calculate the recoil velocity, remnant mass, and spin from the gravitational waveform. Although these formulas differ from the RIT using the isolated horizon algorithm, they carry the same physical meaning.

The energy of gravitational wave radiation $E_{\text{rad}}(t)$ can be calculated from the Weyl scalar $\Psi_4$ \cite{Campanelli:1998jv,Lousto:2007mh}:
\begin{equation}\label{eq:12}
E_{\mathrm{rad}}(t)=\lim _{r \rightarrow \infty} \frac{r^2}{16 \pi} \int_{t_0}^t \mathrm{~d} t^{\prime} \oint_{S_r^2} \mathrm{~d} \Omega\left|\int_{-\infty}^{t^{\prime}} \mathrm{d} t^{\prime \prime} \Psi_4\right|^2,
\end{equation}
where $S_r^2$ represents a space-like slice of null infinity.
Using the orthogonality of ${ }_{-2}Y_{l, m}(\theta, \phi)$ and Eq. (\ref{eq:1}), we can rewrite the radiated energy as
\begin{equation}\label{eq:13}
E_{\mathrm{rad}}(t)=\lim _{r \rightarrow \infty} \frac{r^2}{16 \pi}\sum_{l, m} \int_{t_0}^t \mathrm{~d} t^{\prime}\left|\int_{-\infty}^{t^{\prime}} \mathrm{d} t^{\prime \prime} \Psi_4^{lm}\right|^2.
\end{equation}
Radiated linear momentum $\mathbf{P}_{\mathrm{rad}}(t)$ can be expressed as
\begin{equation}\label{eq:14}
\mathbf{P}_{\mathrm{rad}}(t)=\lim _{r \rightarrow \infty} \frac{r^2}{16 \pi} \int_{t_0}^t \mathrm{~d} t^{\prime} \oint_{S_r^2}\mathrm{~d}\Omega \hat{\mathbf{e}}_r\left|\int_{-\infty}^{t^{\prime}} \mathrm{d} t^{\prime \prime} \Psi_4\right|^2,
\end{equation}
where $\hat{\mathbf{e}}_r$ is the flat space unit radial vector
\begin{equation}\label{eq:15}
\hat{\mathbf{e}}_r=(\sin \theta \cos \phi, \sin \theta \sin \phi, \cos \theta).
\end{equation}
Using the orthogonality of ${ }_{-2}Y_{l, m}(\theta, \phi)$, Eq. (\ref{eq:1}) and the property of the radial unit vector $\hat{\mathbf{e}}_r$, we can rewrite the radiated linear momentum as
\begin{equation}\label{eq:16}
\begin{aligned}
P_{+}^{\mathrm{rad}}  &=\lim _{r \rightarrow \infty} \frac{r^2}{8 \pi} \sum_{l, m} \int_{t_0}^t \mathrm{~d} t^{\prime}\left[\int_{-\infty}^t d t^{\prime \prime} \Psi_4^{l, m}\right. \\
 &\times \int_{-\infty}^t d t^{\prime \prime}\left(a_{l, m} \bar{\Psi}_4^{l, m+1}+b_{l,-m} \bar{\Psi}_4^{l-1, m+1}\right.\\ 
&\left.-b_{l+1, m+1} \bar{\Psi}_4^{l+1, m+1}\right)\bigg],
\end{aligned}
\end{equation}
\begin{equation}\label{eq:17}
\begin{aligned}
P_{z}^{\mathrm{rad}} & =\lim _{r \rightarrow \infty} \frac{r^2}{16 \pi} \sum_{l, m}\int_{t_0}^t \mathrm{~d} t^{\prime} \left[ \int_{-\infty}^t d t^{\prime \prime} \Psi_4^{l, m}\right. \\
& \times \int_{-\infty}^t d t^{\prime \prime}\left(c_{l, m} \bar{\Psi}_4^{l, m}+d_{l, m} \bar{\Psi}_4^{l-1, m}\right. \\
& \left.+d_{l+1, m} \bar{\Psi}_4^{l+1, m}\right)\bigg],
\end{aligned}
\end{equation}
where $P_{+}^{\mathrm{rad}}$ in Eq. (\ref{eq:16}) is a combination quantity introduced for convenience, which is $P_{+}^{\mathrm{rad}}=P_x^{\mathrm{rad}}+\mathrm{i} P_y^{\mathrm{rad}}$. $P_x^{\mathrm{rad}}$, $P_y^{\mathrm{rad}}$ and $P_z^{\mathrm{rad}}$ are the $x$,  $y$ and $z$ components of $\mathbf{P}_{\mathrm{rad}}$ respectively. $\bar{\Psi}_4^{l, m}$ is the conjugate complex of $\Psi_4^{l, m}$. The initial time $t_0$ should exclude the nonphysical radiation relaxation time when specifically calculated. The coefficients $\left(a_{l, m}, b_{l, m}, c_{l, m}, d_{l, m}\right)$ in Eqs. (\ref{eq:16}) and (\ref{eq:17}) are given by
\begin{equation}\label{eq:18}
\begin{aligned}
a_{l, m} &= \frac{\sqrt{(l-m)(l+m+1)}}{l(l+1)}  \\
b_{l, m} &= \frac{1}{2 l} \sqrt{\frac{(l-2)(l+2)(l+m)(l+m-1)}{(2 l-1)(2 l+1)}} \\
c_{l, m} &= \frac{2 m}{l(l+1)}  \\
d_{l, m} &= \frac{1}{l} \sqrt{\frac{(l-2)(l+2)(l-m)(l+m)}{(2 l-1)(2 l+1)}}.
\end{aligned}
\end{equation}
Finally, the radiated angular momentum $\mathbf{J}_{\mathrm{rad}}(t)$ is given by
\begin{equation}\label{eq:19}
\begin{aligned}
\mathbf{J}_{\mathrm{rad}}(t)=&-\lim _{r \rightarrow \infty}  \frac{r^2}{16 \pi} \operatorname{Re} \int_{t_0}^t \mathrm{~d} t^{\prime}\left\{\oint_{S_r^2}\left(\int_{-\infty}^{t^{\prime}} \mathrm{d} t^{\prime \prime} \bar{\Psi}_4\right)\right. \\
& \left.\times \hat{\mathbf{J}}\left(\int_{-\infty}^{t^{\prime}} \mathrm{d} t^{\prime \prime} \int_{-\infty}^{t^{\prime \prime}} \mathrm{d} t^{\prime \prime \prime} \Psi_4\right) \mathrm{d} \Omega\right\},
\end{aligned}
\end{equation}
where the angular momentum operator $\hat{\mathbf{J}}$ for spin weight
$s =-2$ is given by
\begin{equation}\label{eq:20}
\hat{\mathbf{J}}=\left(\operatorname{Re} \hat{\mathbf{J}}_{+}, \operatorname{Im} \hat{\mathbf{J}}_{+}, \frac{\partial}{\partial \phi}\right)
\end{equation}
and
\begin{equation}\label{eq:21}
\hat{\mathbf{J}}_{+}=\mathrm{e}^{\mathrm{i} \phi}\left(\mathrm{i} \frac{\partial}{\partial \theta}-\cot \theta \frac{\partial}{\partial \phi}+2 \mathrm{i} \csc \theta\right).
\end{equation}
Again, using the orthogonality of ${ }_{-2}Y_{l, m}(\theta, \phi)$, Eq. (\ref{eq:1}) and the property of the angular momentum operator $\hat{\mathbf{J}}$, we can rewrite the radiated angular momentum as
\begin{equation}\label{eq:22}
\begin{aligned}
J_{x}^{\mathrm{rad}} & =-\lim _{r \rightarrow \infty} \frac{i r^2}{32 \pi} \operatorname{Im}\left\{\sum_{l, m}\int_{t_0}^t\left[ \int_{-\infty}^{t^{\prime}} \int_{-\infty}^{t^{\prime\prime}} \Psi_4^{l, m} d t^{\prime \prime\prime} d t^{\prime\prime}\right.\right. \\
& \left.\left.\times \int_{-\infty}^{t^{\prime}}\left(f_{l, m} \bar{\Psi}_4^{l, m+1}+f_{l,-m} \bar{\Psi}_4^{l, m-1}\right) d t^{\prime\prime}\right]\mathrm{~d} t^{\prime}\right\},
\end{aligned}
\end{equation}
\begin{equation}\label{eq:23}
\begin{aligned}
J_{y}^{\mathrm{rad}} & =-\lim _{r \rightarrow \infty} \frac{r^2}{32 \pi} \operatorname{Re}\left\{\sum_{l, m} \int_{t_0}^t  \left[\int_{-\infty}^{t^{\prime}} \int_{-\infty}^{t^{\prime\prime}} \Psi_4^{l, m} d t^{\prime \prime\prime} d t^{\prime\prime}\right.\right. \\
& \left.\left.\times \int_{-\infty}^{t^{\prime}}\left(f_{l, m} \bar{\Psi}_4^{l, m+1}-f_{l,-m} \bar{\Psi}_4^{l, m-1}\right)d t^{\prime\prime}\right]\mathrm{~d} t^{\prime}\right\},
\end{aligned}
\end{equation}
\begin{equation}\label{eq:24}
\begin{aligned}
J_{z}^{\mathrm{rad}} & =-\lim _{r \rightarrow \infty} \frac{i r^2}{16 \pi} \operatorname{Im}\left\{\sum_{l, m} m \int_{t_0}^t \left(  \int_{-\infty}^{t^{\prime}}\int_{-\infty}^{t^{\prime\prime}} \Psi_4^{l, m} d t^{\prime\prime\prime}\right.\right. \\
& \left.\left.\times d t^{\prime \prime} \int_{-\infty}^{t^{\prime}} \bar{\Psi}_4^{l, m}d t^{\prime\prime}\right)\mathrm{~d} t^{\prime}\right\},
\end{aligned}
\end{equation}
where the coefficients $f_{l, m}$ in Eqs. (\ref{eq:22}) and (\ref{eq:23}) are given by
\begin{equation}\label{eq:25}
\begin{aligned}
f_{l, m} & :=\sqrt{(l-m)(l+m+1)} \\
& =\sqrt{l(l+1)-m(m+1)}.
\end{aligned}
\end{equation}
The recoil velocity $V_f$ can then be calculated from the radiated linear momentum
\begin{equation}\label{eq:26}
V_f=-\frac{\mathbf{P}_{\mathrm{rad}}}{M_{f}},
\end{equation}
where $M_{f}$ can be calculated from the energy balance:
\begin{equation}\label{eq:27}
M_{f}=M_{\mathrm{ADM}}-E_{\mathrm{rad}}.
\end{equation}
For nonspinning BBH, according to symmetry, the spin direction of the final remnants is in $z$ direction, which can be calculated by
\begin{equation}\label{eq:28}
\alpha_{f}=\frac{L-J_z^{\mathrm{rad}}}{M_{f}^2},
\end{equation}
where $L$ represents the initial orbital angular momentum.

In panel (d) of FIG. \ref{FIG:2} and panel (d) of FIG. \ref{FIG:3}, we observe that before the final peak or valley, the change in $M_f$ is negligible. Therefore, the calculation of dynamic quantities such as $V_f$, $M_f$, and $\alpha_f$ is based mainly on $\mathbf{P}_{\mathrm{rad}}$, $E_{\mathrm{rad}}$, and $J_z^{\mathrm{rad}}$. The calculation of $\mathbf{P}_{\mathrm{rad}}$, $E_{\mathrm{rad}}$, and $J_z^{\mathrm{rad}}$ is based on Eqs. (\ref{eq:13}), (\ref{eq:16}), and (\ref{eq:24}). As we previously mentioned, in the case of a nonspinning BBH, the recoil occurs within the orbital plane, while angular momentum radiation takes place along the $z$-direction. Eqs. (\ref{eq:13}) and (\ref{eq:24}) demonstrate a similarity in the computation of $E_{\mathrm{rad}}$ and $J_z^{\mathrm{rad}}$, except for an additional time integral in $J_z^{\mathrm{rad}}$. This additional integral does not affect the physical regularity of the variation in $J_z^{\mathrm{rad}}$, similar to the oscillation similarity shown in FIGs. \ref{FIG:2} and \ref{FIG:3} between $E_{\mathrm{rad}}$ and $J_z^{\mathrm{rad}}$. However, it introduces a phase shift in $M_f$ relative to the initial eccentricity, which could explain the deviation of the integer cycles $N_{\text {waves }}$ from the peaks and valleys in FIG. \ref{FIG:4}.
Now, let us return to the calculation of the radiated linear momentum (recoil velocity) in Eq. (\ref{eq:16}). If we disregard the last two terms, $b_{l,-m} \bar{\Psi}_4^{l-1, m+1}$ and $b_{l+1, m+1} \bar{\Psi}_4^{l+1, m+1}$, the remaining integral closely resembles Eq. (\ref{eq:13}). The integral over $m$ or the integral over $m+1$ and the coefficient $a_{l, m}$ in Eq. (\ref{eq:16}) do not affect the regularity of the physics. However, including $b_{l,-m} \bar{\Psi}_4^{l-1, m+1}$ and $b_{l+1, m+1} \bar{\Psi}_4^{l+1, m+1}$ introduces significant complexity since it involves the superposition of different harmonic modes, resulting in messy and irregular recoil velocities, as depicted in panel (b) of FIGs. \ref{FIG:2} and \ref{FIG:3}. The irregularities in recoil velocities can be characterized by the following.

(i) The distribution of peaks and valleys in recoil velocities is irregular, without a specific location such as an integer cycle $N_{\text{orbits}}$ of orbital transitions, and they lack uniform sharpness.

(ii) The difference in the cycle numbers $\Delta N_{\text{orbits}}$ between the peaks and valleys is less than 1 and irregular, varying between 0.2 and 0.7.

(iii) The values of the peaks and valleys exhibit irregularity, occasionally causing sudden rises or falls (as seen in panel (b) of FIG. \ref{FIG:2}), and sometimes even surpassing the preceding peak (as seen in panel (b) of FIG. \ref{FIG:3}). This is analogous to the bimodal structure depicted in panel (b) of FIG. \ref{FIG:3}. At first glance, this structure may appear anomalous, but upon understanding the irregular nature of recoil and the coarse-graining resulting from limited simulated data, it becomes apparent why it has such a structure.

These formulas also provide an explanation for the asymptotic behavior of dynamic quantities at high eccentricities and head-on limits. In the scenario after the last valley, $M_f$ gradually increases toward a specific value. Consequently, the recoil velocity $V_f$ and the spin of the remnant $\alpha_f$ in Eqs. (\ref{eq:26}) and (\ref{eq:28}) exhibit a rapid decrease, as observed in panels (b) and (e) of FIG. \ref{FIG:2} and FIG. \ref{FIG:3}. On the other hand, the peak luminosity demonstrates a slow decrease, similar to the behavior of $M_f$, as depicted in panels (c) and (d) of FIG. \ref{FIG:2} and FIG. \ref{FIG:3}.

It is noteworthy to mention this in Ref. \cite{Radia:2021hjs}, a regular functional relationship between recoil velocity and infall direction was acquired, indicating the presence of an intrinsic correlation between these two quantities. However, due to the absence of trajectory information in the RIT catalog, our investigation of the relationship among recoil velocity, eccentricity, and infall direction remains incomplete. We recognize the need for further research in this area to address this limitation.

In summary, the dynamical quantities, including peak luminosity, recoil velocity, mass, and spin of the remnant, display distinct behaviors in their oscillations. Oscillations of peak luminosity, remnant mass, and spin exhibit a more regular pattern, whereas those of recoil velocity appear messy and irregular. This phenomenon can be attributed to their distinct physical origins, specifically the differences in their calculation methods.

Furthermore, we delve into the physical origin of these oscillations, which can be attributed to orbital transitions. As evident from FIG. \ref{FIG:2} and FIG. \ref{FIG:3}, the position of the peaks and valleys in the oscillations corresponds to the appearance of orbital transitions. In other words, these transitions introduce excitations that amplify the dynamic quantities, including peak luminosity, recoil velocity, and the mass and spin of the remnant. This effect manifests itself earlier for small initial coordinate separations, later for larger separations, and typically within less than 10 orbital cycles. Each peak and valley here has the same meaning as the transition from inspiral to plunge. They are close to extreme relativistic situations and only appear at close coordinate separations. So, it is a strong field effect that necessitates NR and cannot be captured by analytical PN methods.
Moreover, we observe that as the eccentricity gradually decreases, the number of orbital cycles increases, resulting in a gradual reduction of the oscillations. This observation provides an alternative perspective, highlighting how orbit averaging can mitigate the impact of eccentricity. However, this effect is only prominent in strong-field regimes and attempts to study it using analytical PN methods can only capture an average in gravitational wave background over a few gravitational wavelengths \cite{Peters:1964zz,Peters:1963ux}. Therefore, the complete manifestation of the orbital averaging effect for eccentricity requires the use of NR.
\begin{figure*}[htbp!]
\centering
\includegraphics[width=15cm,height=20cm]{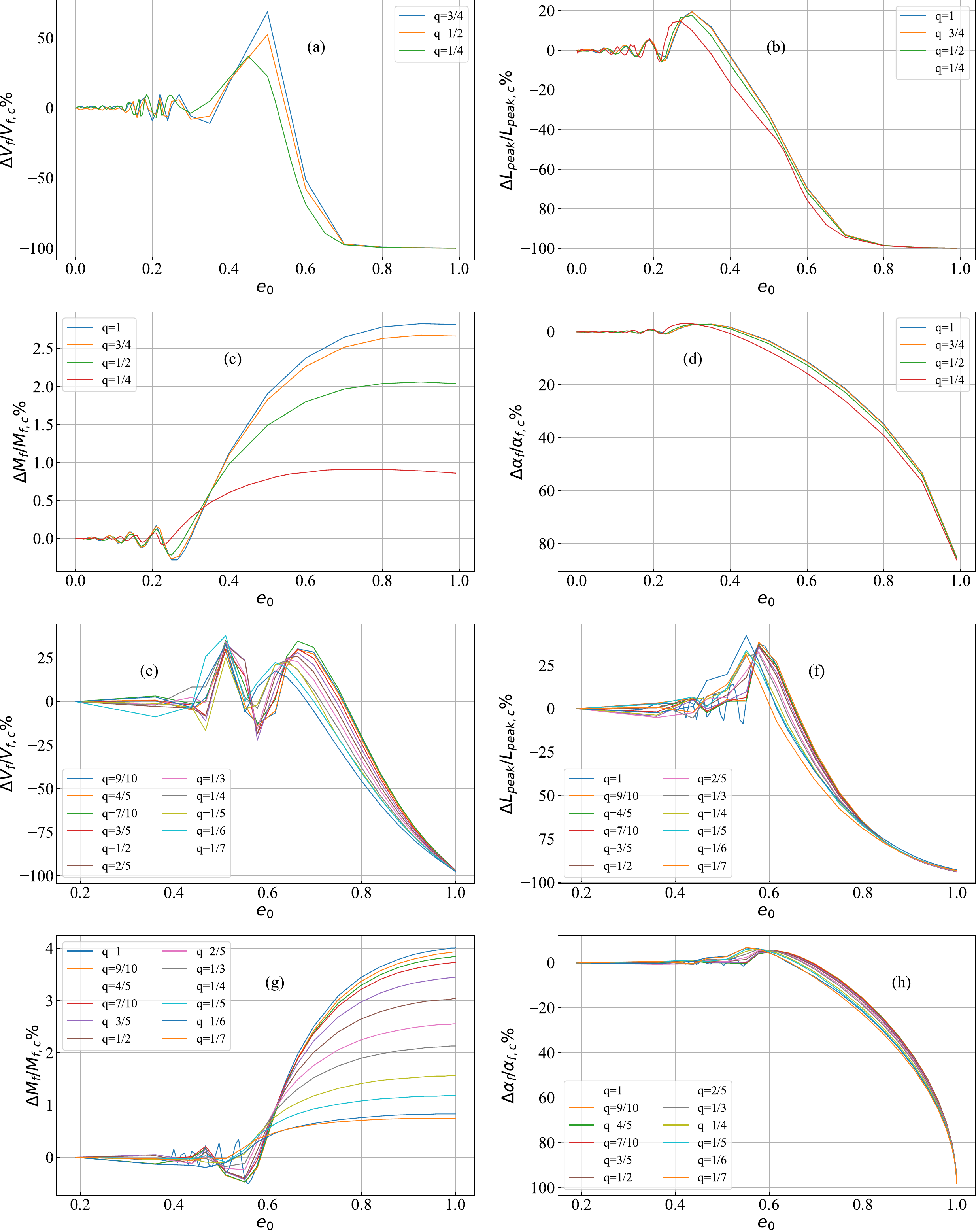}
\caption{\label{FIG:5}Increment percentages of $V_f$, $L_{\text{peak}}$, $M_f$, and $\alpha_f$ relative to the corresponding circular orbit at initial coordinate separations of $11.3M$ and $24.6M$ for nonspinning configuration with different mass ratios. The upper four panels correspond to the initial coordinate separation of $11.3M$, and the lower four panels correspond to the initial coordinate separation of $24.6M$.}
\end{figure*}

To ensure the validity of our research, it is crucial to take into account the errors arising from numerical simulations. Consequently, we address the significance of numerical errors in FIG. \ref{FIG:8} of Appendix \ref{App:A} to confirm that these oscillations are not attributed to numerical errors.

To provide a clearer understanding of the influence of different initial separations $11.3M$ and $24.6M$ on the dynamics of BBH mergers, we present comparisons of dynamic quantities using three eccentricity definitions (RIT, ADM, harmonic) in FIGs. \ref{FIG:13}, \ref{FIG:14}, and \ref{FIG:15} of Appendix \ref{App:C}. This comparison effectively illustrates the impact of the initial coordinate separation on the starting point of the oscillation.
{We also mentioned in Sec. \ref{sec:III} that the comparison of two series of simulations with different initial separations reflects the oscillation behavior of dynamic quantities with eccentricity in the strong field regime. As the initial separation increases, the oscillations manifest at higher initial eccentricities and exhibit more vigorous patterns. This behavior is a direct consequence of the larger initial separation and the increased velocity of the binary black hole prior to merger. A direct comparison of simulations with varying initial distances effectively captures and illustrates this characteristic.   

The influence of initial eccentricity on the enhancement or weakening of dynamic quantities such as peak luminosity, recoil velocity, mass, and spin of the remnant has significant astrophysical implications. From a PN perspective, Ref. \cite{Sopuerta:2006et} suggests a proportional relationship between recoil velocity ($V_f$) and low eccentricity ($e_0$), i.e., $V_f \propto (1+e_0)$. However, in the strong-field regime of NR, no obvious proportional relationship between recoil and initial eccentricity has been observed in FIG. \ref{FIG:2} and FIG. \ref{FIG:3}. Previous studies, such as Refs. \cite{Radia:2021hjs} and \cite{Sperhake:2019wwo}, have quantitatively analyzed the enhancement effect induced by eccentricity.
To quantitatively analyze the relative increment percentage of peak luminosity ($L_{\text{peak}}$), recoil velocity ($V_f$), mass ($M_f$), and spin ($\alpha_f$) of the remnant relative to the corresponding circular orbit, we express it as:
\begin{equation}\label{eq:29}
\frac{\Delta A}{A_c}=\frac{A_e-A_c}{A_c} \times 100\%,
\end{equation}
where $A$ denotes $L_{\text{peak}}$, $V_f$, $M_f$, or $\alpha_f$, and the subscripts $e$ and $c$ represent the cases of eccentric and corresponding circular orbits, respectively.
For the initial coordinate separation of $11.3M$, the first set of simulations with zero eccentricity serves as $A_c$. However, for the initial coordinate separation of $24.6M$, despite an initial eccentricity of 0.19 in the first group, the near-horizontal characteristic observed in FIG. \ref{FIG:3} makes it comparable to a circular orbit, allowing us to approximate it as $A_c$. It should be noted that we exclude recoil with a mass ratio of $q=1$ and the cases with mass ratios of $q=1/6$ and $q=1/32$ at the initial coordinate separation of $24.6M$ due to the unreasonable initial eccentricity ($e_0=0.51$) to approximate a circular orbit.
In FIG. \ref{FIG:5}, we present the percentages of increase of $L_{\text{peak}}$, $V_f$, $M_f$, and $\alpha_f$ relative to the corresponding circular orbit for the initial coordinate separations of $11.3M$ and $24.6M$. Notable observations include:

(i) The relative increase of the dynamic quantities is influenced by the initial coordinate separation and the mass ratio. Here we focus solely on peaks or valleys and exclude discussions on high eccentricity and the head-on collision limit. We find that for the recoil velocity $V_f$, at the initial coordinate separation of $11.3M$, the maximum relative increase can reach 69\% for $q=3/4$, while at $24.6M$, the maximum relative increase is 38\% for $q=1/6$. There is a remarkably significant increase in their values compared to circular orbits. As for peak luminosity $L_{\text{peak}}$, at the initial coordinate separation of $11.3M$, the maximum relative increment can reach 20\% for $q=3/4$, and at $24.6M$, the maximum relative increment is 42\% for $q=1/6$. Regarding mass $M_f$, at the initial coordinate separation of $11.3M$, the minimum relative increment can reach -0.28\% for $q=1$, and at $24.6M$, the minimum relative increment is -0.5\% for $q=1$. Lastly, for spin $\alpha_f$, at the initial coordinate separation of $11.3M$, the maximum relative increase can reach 3.1\% for $q=1/4$, while at $24.6M$, the maximum relative increase is 6.9\% for $q=1/7$.

(ii) In the case of regular oscillations ($L_{\text{peak}}$, $M_f$, $\alpha_f$), the last orbital transition from orbit to plunge introduces the most significant relative increment or decrement, leading to a substantial change compared to the penultimate peak or valley. This observation is evident in panels (b), (c), (d), (f), (g), and (h) of FIG. \ref{FIG:5}.
For further details and additional features, that are not discussed comprehensively here, please refer to FIG. \ref{FIG:5}.

One may initially attribute these oscillations to fractional orbital effects, specifically varying fractional orbital numbers. However, it is important to note that the orbital transition effect induced by eccentricity represents a complete strong field phenomenon. This effect alters the characteristics of the merging binary black holes (BBHs), most notably the peak luminosity, during the merger stages of plunge and merger, which are the primary contributors to gravitational radiation. As the number of orbits increases, the properties of eccentric mergers gradually converge towards those of circular orbit mergers. This convergence is evident in the similarity of recoil velocities, peak luminosities, masses, and spins observed at the initial points of the curves in FIG. \ref{FIG:2} (The recoil speeds at mass ratios of 1/2 and 1/4 exhibit deviations from other initial points due to numerical errors. However, the impact of these deviations is still within the range of recoil enhancement caused by eccentricity). The properties of BBH mergers for various circular orbits are denoted by black marks ``x". These circular orbits correspond to RIT:BBH:0001 ($q=1$), RIT:BBH:0112 ($q=1$), RIT:BBH:0198 ($q=1$), RIT:BBH:0114 ($q=3/4$), RIT:BBH:0117 ($q=1/2$), and RIT:BBH:0119 ($q=1/4$) in the RIT catalog, respectively. Their initial orbital separations are 9.53, 20.0, 11.0, 11.0, 11.0, and 11.0, respectively. While these different initial separations result in varying fractional orbits, the influence of fractional orbits is considerably less significant compared to the impact of the orbital transition effect caused by eccentricity.

\subsubsection{Summary}\label{sec:III:A:4}
In conclusion, in Section \ref{sec:III:A}, we have provided a comprehensive analysis of the relationship between various dynamic quantities, including merger time $T_{\text{merger}}$, peak luminosity $L_{\text{peak}}$, recoil velocity $V_f$, mass $M_f$, and spin $\alpha_f$ of the merger remnants, and the initial eccentricity $e_0$ for different initial coordinate separations $11.3M$ and $24.6M$. Our findings reveal intriguing oscillatory behaviors, which become evident when the numerical simulation data points are sufficiently dense.
In Sec. \ref{sec:III:A:1} and Sec. \ref{sec:III:A:2}, we objectively described the observed phenomenology and oscillations without delving into their physical origins. However, in Sec. \ref{sec:III:A:3}, we embarked on exploring the underlying causes of these oscillations. Using the phase of gravitational waves, we calculated the orbital cycle number $N_{\text{orbits}}$ and found a remarkable correlation between peaks or valleys of the dynamic quantities and orbital transitions. Subsequently, we employed calculation formulas analysis from gravitational waveform to examine the oscillatory behavior exhibited by different dynamic quantities. Our analysis led us to conclude that the distinct oscillation patterns observed in various physical quantities arise from the use of different calculation methods.
Finally, to address the astrophysical implications of our findings, we computed the percentage increment of each dynamic quantity in eccentric orbits relative to corresponding circular orbits. This analysis provides valuable insights into the relative enhancements or weakenings of these quantities associated with eccentricity.
In general, our study sheds light on the intricate relationship between initial eccentricity and dynamic quantities, revealing oscillatory phenomena and providing a deeper understanding of their physical origins. The calculated percentage increments further contribute to our understanding of the astrophysical implications of eccentric orbits compared to circular orbits.

\subsection{Spin alignment}\label{sec:III:B}
\subsubsection{Analysis}
\begin{table*}[htbp!]
\caption{\label{tab:I}Parameters for eccentric BBH simulations with spin-aligned configurations, where $e_{0,min}$ represents the minimum value of the initial eccentricity in the simulation series.}
    \centering
    \begin{tabular}{|r|r|r|r|r|r|r|r|}
    \hline
        configuration ID & $q$ & $\chi_{1z}$ & $\chi_{2z}$ & $\chi_{\mathrm{eff}}$ & $e_{0,min}$ & set number  \\ \hline
        A1 & 1 & -0.5 & -0.5 & -0.5 & 0.19 & 23   \\ \hline
        A2 & 1 & -0.8 & -0.8 & -0.8 & 0.19 & 23 \\ \hline
        A3 & 1 & 0.5 & 0.5 & 0.5 & 0.4375 & 21  \\ \hline
        A4 & 1 & 0.8 & 0.8 & 0.8 & 0.4375 & 21  \\ \hline
        A5 & 1 & 0 & 0.8 & 0.4 & 0.4375 & 14   \\ \hline
        A6 & 1 & 0 & -0.8 & -0.4 & 0.19 & 14   \\ \hline
        A7 & 1/4 & 0 & -0.8 & -0.64 & 0.4375 & 20   \\ \hline
        A8 & 1/3 & 0 & -0.8 & -0.6 & 0.36 & 21   \\ \hline
        A9 & 1/2 & 0 & -0.8 & -0.53 & 0.36 & 16  \\ \hline
    \end{tabular}
\end{table*}
\begin{figure*}[htbp!]
\centering
\includegraphics[width=15cm,height=15cm]{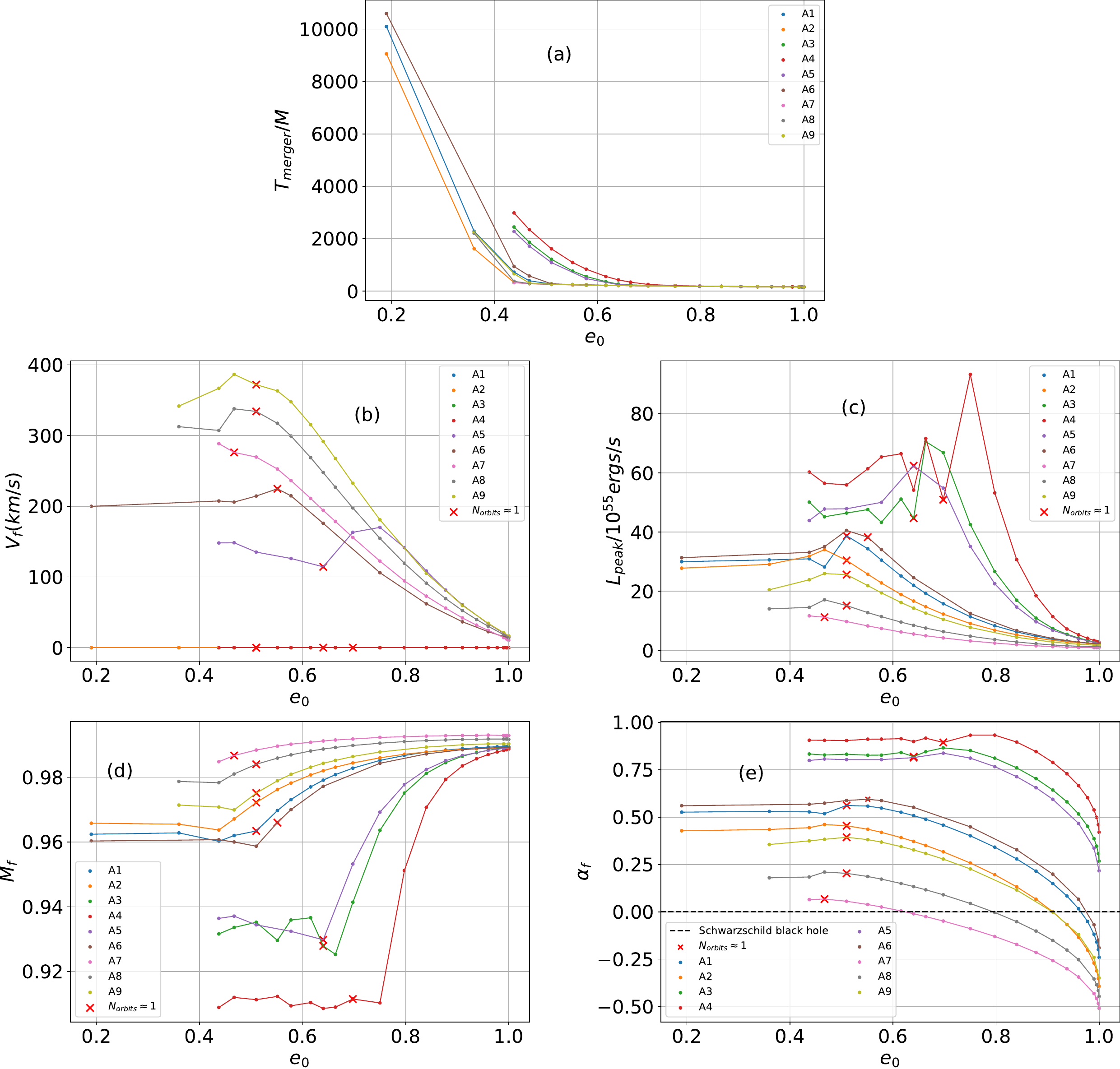}
\caption{\label{FIG:6}Variations of dynamical quantities of the merger time $T_{\text{merger}}$ (panel (a)), peak luminosity $L_{\text{peak}}$ (panel (c)), recoil velocity $V_f$ (panel (b)), mass $M_f$ (panel (d)), and spin $\alpha_f$ (panel (e)) of the merger remnants as a function of the initial eccentricity $e_0$ at the initial coordinate separation of $24.6M$ for spin aligned configuration with different mass ratios. We mark the position where $N_{\text{orbits}}$ is approximately equal to one orbit with a red ``x". The dashed line in panel (e) represents the Schwarzschild black hole whose corresponding spin is 0.}
\end{figure*}
The analysis of spin-aligned eccentric BBH mergers follows a similar framework to the previous nonspinning case. However, the inclusion of spin introduces additional considerations. Specifically, we need to account for the influence of spin on the merger dynamics. The hangup effect, characterized by spin alignment or anti-alignment with orbital angular momentum, can either slow down or accelerate the BBH merger compared to the nonspinning scenario \cite{Lousto:2011kp,Lousto:2012su,Hannam:2007wf,Campanelli:2006uy}. This effect fundamentally alters the relationship between the dynamic quantities of the BBH merger, including the merger time $T_{\text{merger}}$, the peak luminosity $L_{\text{peak}}$, the recoil velocity $V_f$, the mass $M_f$, and the spin $\alpha_f$, with respect to the initial eccentricity $e_0$, relative to the nonspinning case.
TABLE \ref{tab:I} provides the parameters for eccentric BBH simulations with spin-aligned or anti-aligned configurations (collectively referred to as spin-aligned for simplicity) from RIT \cite{RITBBH}. It is important to note that the minimum values ${e_0}_{\text{min}}$ of the initial eccentricity differ across the simulations, and the maximum value of the initial eccentricity is set to 0.9999, approaching the head-on collision limit. To facilitate representation and analysis, each simulation configuration is assigned a unique ID, as indicated in the first column of TABLE \ref{tab:I}.  FIG. \ref{FIG:6} illustrates the dynamic quantities merger time $T_{\text{merger}}$, peak luminosity $L_{\text{peak}}$, recoil velocity $V_f$, mass $M_f$, and spin $\alpha_f$, as functions of the initial eccentricity $e_0$ for the spin-aligned configuration at an initial coordinate separation of $24.6M$. Additionally, we mark the position where $N_{\text{orbits}}$ is approximately equal to one orbit with a red ``x" in FIG. \ref{FIG:6}. Notably, we now consider different effective spin configurations, which introduce variations compared to the nonspinning BBH case.
Incorporating spin into the analysis of eccentric BBH mergers enhances our understanding of the complex interaction between spin dynamics and initial eccentricity. The inclusion of different effective spin configurations further enriches the investigation of the orbital hangup effect, highlighting nuances compared to the nonspinning case.

In panel (a) of FIG. \ref{FIG:6}, the relationship between merger time and initial eccentricity exhibits similarities to the overall behavior observed in the previous nonspinning BBH case, as depicted in FIG. \ref{FIG:2} and FIG. \ref{FIG:3}. However, the hangup effect significantly alters the location of the critical point in the merger time. Specifically, for positive effective spin values, the corresponding initial eccentricity at the critical point is higher, approximately 0.65. On the contrary, for negative effective spin values, the critical point occurs at a lower initial eccentricity, around 0.45. This observation underscores the profound impact of the hangup effect on either accelerating or decelerating the BBH merger process. In particular, a greater effective spin leads to longer merger times, indicating a stronger influence of spin on the dynamics of the system.

In panel (b) of FIG. \ref{FIG:6}, the overall behavior of the recoil velocity aligns with the trends observed in FIG. \ref{FIG:2} and FIG. \ref{FIG:3}. However, because of limited data points and a scarcity of simulations with low initial eccentricity, the oscillatory behavior is not clearly visible, and the final peak is barely discernible. When the mass ratio $q=1$ and the spins are equal, the BBH system adopts a perfectly symmetric configuration, resulting in zero radiated linear momentum and, consequently, a recoil velocity of 0. Comparing configurations A1, A2, A3, and A4, we observe that the influence of different mass ratios on the recoil velocity persists, similar to the nonspinning case. However, the effect of aligned spin on the recoil velocity is twice as significant as the effect of an asymmetric mass ratio. Previously, the maximum recoil value introduced by an asymmetric mass ratio of $q=1/3$ was 226 km/s, but configuration A9 raises this value to 387 km/s. As the spin configuration becomes more asymmetric, the resulting recoil velocity increases. Simultaneously, the hangup effect causes a shift in the initial eccentricity corresponding to the recoil peak, reflecting its role in accelerating or decelerating the BBH dynamics. Notably, the presence of spin amplifies the recoil velocity for circular orbital cases. Consequently, the incremental percentage of recoil is reduced in the presence of spin compared to the previous nonspinning scenario.

In panel (c) of FIG. \ref{FIG:6}, the overall behavior of the peak luminosity shows similarities to FIGs. \ref{FIG:2} and \ref{FIG:3}. Some configurations, such as A3 and A4, display slight oscillations, consistent with the previous observations. However, it is important to note that these oscillations are not comprehensive, as the available data points are limited and represent a coarse-grained picture.
The influence of spin on the peak luminosity is significantly greater than the effect of eccentricity. For the simulation sequences in RIT, in the absence of spin and eccentricity, the maximum value of peak luminosity can reach $5.1 \times 10^{56}$ ergs/s. However, with spin and no eccentricity, the maximum value of peak luminosity can reach $7.0 \times 10^{56}$ ergs/s. When both eccentricity and spin are present, as in configuration A4, the maximum value of peak luminosity can reach $9.3 \times 10^{56}$ ergs/s.
In panel (c), the impact of the hangup effect on the peak luminosity is also evident, which will not be detailed here.
The orbital cycle number $N_{\text {orbits }}\approx1$ is approximately located near the peak, similar to the situation without spin. However, due to the limited data points and inherent uncertainties, this value should be regarded as a reference rather than an exact measurement.

The analysis of panel (d) in FIG. \ref{FIG:6} follows a similar pattern to the previous panels (b) and (c), and therefore, we will refrain from repeating it here.

In panel (e) of FIG. \ref{FIG:6}, the overall behavior of the spin of the remnant exhibits similarities to FIGs. \ref{FIG:2} and \ref{FIG:3}. However, the presence of spin introduces a new phenomenon in the presence of eccentricity i.e. a final spin transition from positive to negative, passing through the Schwarzschild black hole during the process. In eccentric BBH simulations, the increase in initial eccentricity is equivalent to the decrease in tangential linear momentum, as can be observed from the relationships $p_t=p_{t, q c}(1-\epsilon)$ and $e=2\epsilon-\epsilon^2$. The initial angular momentum $L$ of the BBH can be expressed as
\begin{equation}\label{eq:30}
L=p_t D.
\end{equation}
In the spin-aligned configuration, the radiated angular momentum $J_z^{\mathrm{rad}}(e_0,q,\chi_{1z},\chi_{2z})$ is in the $z$ direction and depends on the mass ratio $q$, the initial eccentricity $e_0$, and the spins $\chi_{1z}$ and $\chi_{2z}$. Referring to previous work \cite{Sperhake:2007gu,Buonanno:2007sv}, neglecting effects such as high-order spin-orbit coupling and spin-spin coupling, and assuming that the spin of each black hole remains constant during the evolution of the BBH, the approximate expression for the final spin parameters $\alpha_{f}$ is given by 
\begin{equation}\label{eq:31}
\alpha_{f}=\frac{L({e_0}_,q)-J_z^{\mathrm{rad}}(e_0,q,\chi_{1z},\chi_{2z})}{M_{f}^2(e_0,q,\chi_{1z},\chi_{2z})}+\chi_{1z}+\chi_{2z}.
\end{equation}
As previously analyzed, when the initial coordinate separation is fixed, both $L$ and $M_f$ are functions of the initial eccentricity $e_0$ and mass ratio $q$, with the latter also dependent on the spins $\chi_{1z}$ and $\chi_{2z}$. If the spin direction $\chi_{1z}$ and $\chi_{2z}$ aligns with the orbital angular momentum or the sum of $\chi_{1z}$ and $\chi_{2z}$ is greater than 0, regardless of the adjustment of the initial eccentricity $e_0$, the final spin direction remains positive (in accordance with the direction of the orbital angular momentum). On the other hand, if the spin direction $\chi_{1z}$ and $\chi_{2z}$ is anti-aligned with the orbital angular momentum or the absolute value of the sum (This sum is required to be negative) of $\chi_{1z}$ and $\chi_{2z}$ is greater than the first term of the right side of Eq. (\ref{eq:31}), it is possible to finely adjust the initial eccentricity such that the final spin $\alpha_{f}$ becomes 0, resulting in a Schwarzschild black hole. This relationship can be qualitatively expressed as:
\begin{equation}\label{eq:32}
0=\frac{L({e_0}_S,q)-J_z^{\mathrm{rad}}({e_0}_S,q,\chi_{1z},\chi_{2z})}{M_{f}^2({e_0}_S,q,\chi_{1z},\chi_{2z})}+\chi_{1z}+\chi_{2z}.
\end{equation}
Determining accurately the initial eccentricity ${e_0}_S$ that leads to the final black hole being a Schwarzschild black hole is challenging when using analytical modeling. This difficulty arises from the need to consider the eccentricity's special effects as well as the influence of the hangup effect, which makes the problem highly complex. From panel (e) in FIG. \ref{FIG:6}, it can be observed that the initial eccentricities that eventually result in a Schwarzschild black hole are all in the plunge stage rather than in the inspiral stage, indicating high eccentricity and complex strong field dynamics. The corresponding initial eccentricity values to form a Schwarzschild black hole for configurations A1, A2, A6, A7, A8, and A9 are 0.96, 0.91, 0.96, 0.6156, 0.7975, and 0.91, respectively. These eccentricity values do not imply that the final black hole spin is exactly 0, but rather that it is as close as possible to 0. These initial eccentricity values provide insights into the influence of spin and mass ratios on the BBH dynamics. Importantly, it is worth noting that the combined effect of eccentricity and spin does not cause the final black hole's spin to exceed that of an extreme Kerr black hole whose spin is 1, thus confirming the validity of the cosmic censorship hypothesis \cite{Penrose:1964wq,Penrose:1969pc}.

\subsubsection{Summary}
In summary, in Sec. \ref{sec:III:B}, we presented a comprehensive analysis of various eccentric spin alignment configurations in the BBH merger simulations. We investigated the relationship between key dynamic quantities, including the merger time $T_{\text{merger}}$, peak luminosity $L_{\text{peak}}$, recoil velocity $V_f$, mass $M_f$, and spin $\alpha_f$ of the merger remnants, and the initial eccentricity $e_0$ for an initial coordinate separation of $24.6M$. Our findings demonstrate that the overall behavior of these dynamic quantities follows a similar pattern to the nonspinning case. They start with a horizontal line, gradually exhibit oscillations towards the final peak or valley (although due to limited data points, we observed only a portion of the oscillation), and eventually converge to a certain value as they approach the head-on collision limit.
This universal behavior reveals the similar effects of eccentricity on dynamics, regardless of spin alignment or no spin. In both the nonspinning and spin-aligned scenarios, the percentage increment of these dynamic quantities due to eccentricity, relative to the circular orbit case, remains approximately analogous. This observation underscores the universality of eccentricity's influence on BBH dynamics.
The hangup effect plays a crucial role in altering the critical points of the merger time $T_{\text{merger}}$, modifying the baseline value of the recoil velocity and the corresponding eccentricity at the final peak for $V_f$, and introducing variations in the peak luminosity $L_{\text{peak}}$ and remnant mass $M_f$ compared to the case of zero eccentricity. Additionally, it can give rise to a critical eccentricity that results in a transition across the Schwarzschild black hole for $\alpha_f$. These effects, characterized by the alterations in dynamic quantities of BBHs under the influence of spin and eccentricity, have profound astrophysical implications.
\subsection{Spin precession}\label{sec:III:C}
\subsubsection{Analysis}
\begin{table*}[htbp!]
\caption{\label{tab:II}Parameters for eccentric BBH simulations with spin precession configurations, where $e_{0,min}$ represents the minimum value of the initial eccentricity in the simulation series.}
    \centering
    \begin{tabular}{|r|r|r|r|r|r|r|r|r|r|r|}
    \hline
        configuration ID & $q$ & $\chi_{1x}$ & $\chi_{1y}$ & $\chi_{1z}$ & $\chi_{2x}$ & $\chi_{2y}$ & $\chi_{2z}$ & $\chi_{p}$ & $e_{0,min}$ & set number  \\ \hline
        P1 & 1 & 0 & -0.6062 & 0.35 & 0 & 0.6062 & 0.35 & 0.6062 & 0.51 & 7  \\ \hline
        P2 & 1 & 0 & -0.7 & 0 & 0 & 0.7 & 0 & 0.7 & 0.51 & 7  \\ \hline
        P3 & 1 & 0.5 & 0 & 0 & 0.5 & 0 & 0 & 0.5 & 0.5775 & 5  \\ \hline
        P4 & 1 & 0.6 & 0 & 0 & 0.6 & 0 & 0 & 0.6 & 0.5775 & 5  \\ \hline
        P5 & 1 & 0.7 & 0 & 0 & 0.7 & 0 & 0 & 0.7 & 0.19 & 15  \\ \hline
        P6 & 1 & 0.6062 & 0.35 & 0 & 0.7 & 0 & 0 & 0.7 & 0.51 & 11  \\ \hline
        P7 & 1 & 0.35 & 0.6062 & 0 & 0.7 & 0 & 0 & 0.7 & 0.51 & 11  \\ \hline
        P8 & 1 & 0 & 0.7 & 0 & 0.7 & 0 & 0 & 0.7 & 0.51 & 11  \\ \hline
        P9 & 1 & -0.35 & 0.6062 & 0 & 0.7 & 0 & 0 & 0.7 & 0.51 & 11  \\ \hline
        P10 & 1 & -0.6062 & 0.35 & 0 & 0.7 & 0 & 0 & 0.7 & 0.51 & 11  \\ \hline
        P11 & 1 & -0.7 & 0 & 0 & 0.7 & 0 & 0 & 0.7 & 0.19 & 15 \\ \hline
    \end{tabular}
\end{table*}
\begin{figure*}[htbp!]
\centering
\includegraphics[width=15cm,height=15cm]{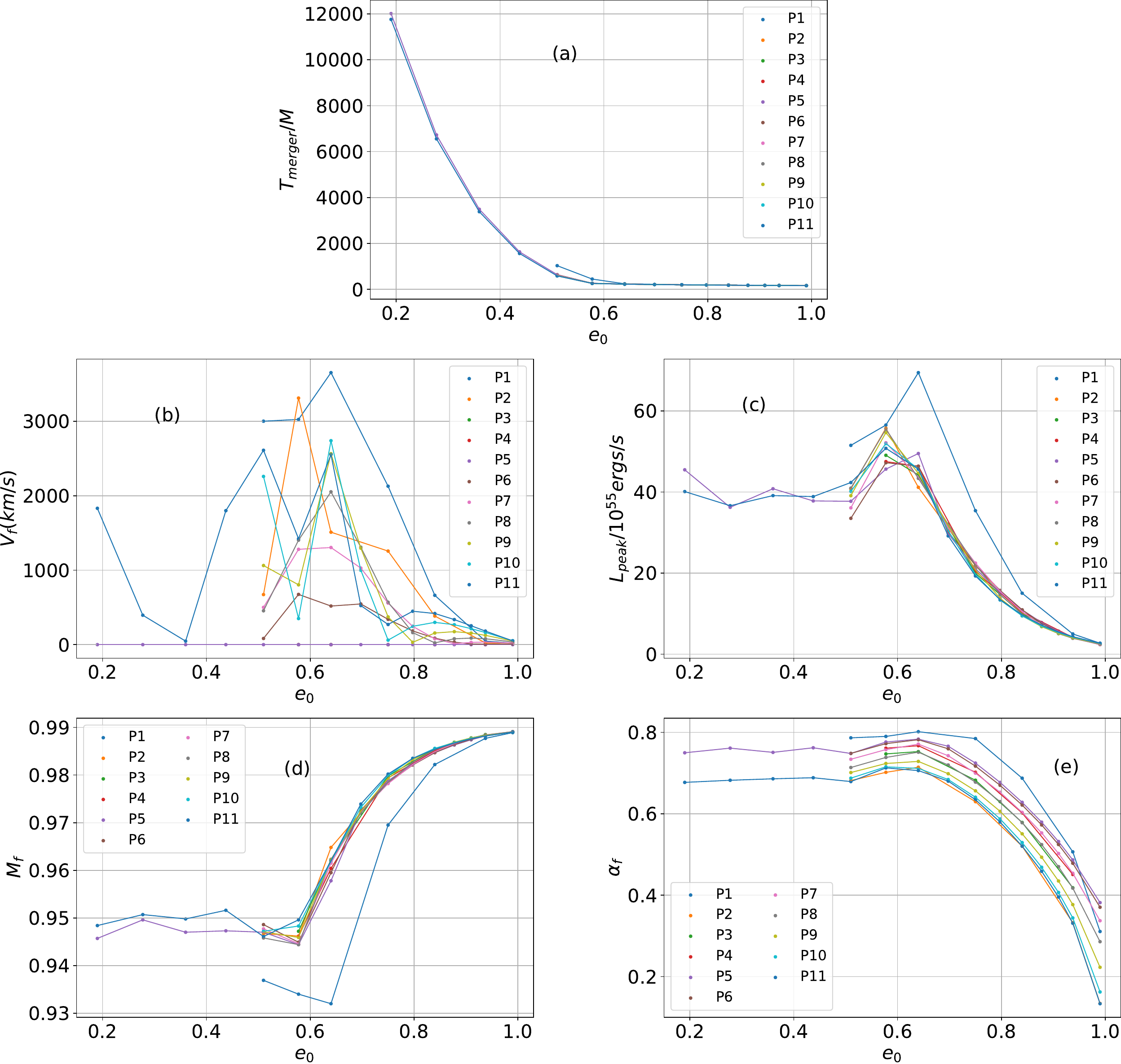}
\caption{\label{FIG:7}Variations of dynamical quantities of the merger time $T_{\text{merger}}$ (panel (a)), peak luminosity $L_{\text{peak}}$ (panel (c)), recoil velocity $V_f$ (panel (b)), mass $M_f$ (panel (d)), and spin $\alpha_f$ (panel (e)) of the merger remnants as a function of the initial eccentricity $e_0$ at the initial coordinate separation of $24.6M$ for spin precession configuration with mass ratio $q=1$.}
\end{figure*}
When the spin angular momentum and orbital angular momentum directions are misaligned, orbital precession can occur. This precession effect introduces intricate modulations on waveforms and dynamics, including amplitude and phase modulation of the waveform and orbital plane precession \cite{Schmidt:2012rh,Apostolatos:1994mx}. The situation becomes even more complex when eccentricity is introduced \cite{Wang:2023ueg}. In this scenario, the waveform undergoes dual modulation. As discussed in Sec. \ref{sec:III:A}, this impact on the waveform is equivalent to the impact on dynamic quantities such as $L_{\text{peak}}$, $V_f$, $M_f$, and $\alpha_f$. Furthermore, precession affects the merger time $T_{\text{merger}}$. In TABLE \ref{tab:II}, we provide the parameters of the eccentric BBH simulations used for spin precession.
In FIG. \ref{FIG:7}, we present the dynamic quantities of the merger remnants, including merger time $T_{\text{merger}}$, peak luminosity $L_{\text{peak}}$, recoil velocity $V_f$, mass $M_f$, and spin $\alpha_f$, as functions of the initial eccentricity $e_0$ for the spin-precessing configuration at an initial coordinate separation of $24.6M$. In FIG. \ref{FIG:7}, we do not mark the position where $N_{\text{orbits}}$ is approximately one orbit due to the limited number of data points. In such cases, the cycle number $N_{\text{orbits}}$ deviates significantly from an integer value and lacks a reference value. To facilitate comparison with the effective spin previously studied, we introduce the effective precession spin parameter $\chi_p$ in an attempt to quantitatively describe the impact of precession. Due to the vast parameter space, RIT's simulations only cover eccentric precession configurations with a mass ratio of $q=1$ and some special spin configurations.

In panel (a) of FIG. \ref{FIG:7}, we observe that the variation of merger time with initial eccentricity exhibits a similar trend to the overall behavior of the nonspinning BBH case in FIG. \ref{FIG:2} and FIG. \ref{FIG:3}, as well as the spin-aligned BBH case in FIG. \ref{FIG:6}. Notably, there are no apparent differences among the effects of different spin precession configurations. The impact is not as pronounced as the changes induced by the hang-up effect observed previously. The critical turning point of $T_{\text{merger}}$ aligns closely with the non-spinning case, occurring at approximately 0.5. Moreover, the effective precession spin parameter exhibits comparable values in all configurations. However, due to the limited number of data points, it is challenging to discern any significant correlations.

In panel (b) of FIG. \ref{FIG:7}, we observe that the variation of the recoil velocity with initial eccentricity follows a trend similar to the overall behavior of the non-spinning BBH case in FIG. \ref{FIG:2} and FIG. \ref{FIG:3}, as well as the spin-aligned BBH case in FIG. \ref{FIG:6}. However, the oscillations in panel (b) appear more chaotic compared to both the nonspinning and spin-aligned cases. It is worth noting that certain configurations, such as P4 and P5, exhibit a recoil velocity of 0 due to symmetry in mass ratio and spin.
Firstly, we observe that the magnitude of the recoil velocity is approximately an order of magnitude larger than in the previous nonspinning and spin-aligned cases. This phenomenon can be attributed to the increased asymmetry exhibited by the precession configuration in comparison to the spin alignment and no spin, leading to higher recoil velocities. Among the configurations, P1 reaches a maximum recoil velocity of 3653.64 km/s at an eccentricity of 0.64, while the smallest cases such as P6 reach a maximum recoil velocity of 674.81 km/s at an eccentricity of 0.5775.
Second, the initial eccentricities at which the maximum recoil values occur for each configuration are not consistent, contributing to the visual complexity in panel (b). As discussed in Sec. \ref{sec:III:A}, we already understand the origin of these chaotic oscillations in the recoil velocity. The messy appearance in panel (b) is a combined effect of eccentricity and spin precession, with spin playing a more dominant role in the observed behavior compared to eccentricity.
Furthermore, we can observe from P11 that the maximum recoil caused by eccentricity is 722 km/s larger than the recoil observed in the corresponding circular orbit. This difference corresponds to a maximum percentage increase of 25.5\%, which is consistent with the findings of previous cases without spin and spin alignment. This quantitative concept holds significant astrophysical significance and provides valuable insights into the dynamics of eccentric BBH systems.

In panels (c), (d), and (e) of FIG. \ref{FIG:7}, we observe that the variations of the peak luminosity $L_{\text{peak}}$, mass $M_f$, and spin $\alpha_f$ of merger remnants with initial eccentricity follow a similar trend to the overall behavior observed in the nonspinning BBH case in FIG. \ref{FIG:2} and FIG. \ref{FIG:3}, as well as the spin-aligned BBH case in FIG. \ref{FIG:6}.
From configurations P5 and P11, we can see that in the presence of spin precession, the incremental percentages of the dynamic quantities $L_{\text{peak}}$, $M_f$, and $\alpha_f$ relative to the values in a circular orbit are essentially consistent with the findings in the no spin and spin-aligned case. These observations indicate that, regardless of the inclusion of spin, the effect of eccentricity on the dynamics of BBHs remains universal and does not change. Furthermore, 
these findings highlight the fact that eccentricity exerts a consistent influence on BBH dynamics, regardless of the presence or absence of spin. They underscore the universal nature of the eccentricity-induced effects and provide further insight into the behavior of eccentric BBH systems. The other detailed analysis is the same as the previous no spin and spin alignment, so we will not go into details here (refer to FIG. \ref{FIG:7}).

\subsubsection{Summary}
In summary, Sec. \ref{sec:III:C} presents a collection of representative simulations of eccentric spin precession configurations in BBH systems. We investigate the relationship between several dynamic quantities, the merger time $T_{\text{merger}}$, peak luminosity $L_{\text{peak}}$, recoil velocity $V_f$, mass $M_f$, and spin $\alpha_f$ of the merger remnants, and the initial eccentricity $e_0$ for an initial coordinate separation of $24.6M$. Our analysis reveals that the overall behavior of these dynamic quantities closely resembles that observed in previous studies involving nonspinning and spin-aligned cases.
However, it is important to note that, due to limitations in the available data points, we do not observe oscillatory patterns similar to those depicted in FIGs. \ref{FIG:2} and \ref{FIG:3}. We conduct an analysis to understand the reasons behind the intricate nature of recoil in panel (b), as illustrated in FIG. \ref{FIG:7}, and propose that it arises from the combined effects of spin precession and eccentricity. Notably, we find that in the presence of spin precession, the percentage increment of the dynamic quantities with respect to the initial eccentricity remains consistent with that observed in both the no spin and spin-aligned scenarios. These findings highlight the universality of the influence of eccentricity on BBH dynamics, which has significant astrophysical implications.

\section{Conclusion and Outlook}\label{sec:IV}
Thanks to the extensive collection of numerical relativistic simulations of eccentric orbital BBH mergers conducted by RIT, we investigated the effect of the initial eccentricity $e_0$ on various dynamic quantities, including the merger time $T_{\text{merger}}$, peak luminosity $L_{\text{peak}}$, recoil velocity $V_f$, mass $M_f$, and spin $\alpha_f$ of the merger remnants. Our study encompasses configurations involving no spin, spin alignment, and spin precession, as well as a wide parameter space that encompasses mass ratios ranging from 1/32 to 1 and initial eccentricities spanning from 0 to 1.

In the case of non-spinning BBH systems, we conducted a detailed investigation using two fixed initial coordinate separations $11.3M$ and $24.6M$. For the $11.3M$ separation, we make a significant discovery regarding the presence of a widespread oscillation phenomenon in the relationship between dynamic quantities $L_{\text{peak}}$, $V_f$, $M_f$, $\alpha_f$, and the initial eccentricity $e_0$. This observation represents the first identification of such universal oscillations in this context. Furthermore, in the case of a mass ratio of $q=1$ and the $24.6M$ separation, we also observe similar oscillatory behavior, leading us to conclude that this phenomenon will manifest itself in numerical simulations featuring sufficiently dense initial eccentricity. We further analyze the role played by the mass ratio in these oscillations.
To gain further insight into these oscillations, we calculate the orbital cycle number $N_{\text{orbits}}$ by examining the phase of gravitational waves. We establish a connection between the integer value of $N_{\text{orbits}}$ and the peaks and valleys observed in the curves of the dynamic quantities. This association leads us to infer that the oscillation phenomenon arises from orbital transitions. This study presents a groundbreaking discovery of the dynamic effects arising from additional orbital transitions in eccentric BBH mergers, beyond the well-known transition from inspiral to plunge \cite{Sperhake:2007gu}. Subsequently, we analyze the formulas used to calculate $V_f$, $M_f$, and $\alpha_f$ from the gravitational waveform. We propose that the chaotic behavior observed in the recoil velocity $V_f$ and the regular behavior observed in $M_f$ and $\alpha_f$ are the result of differences in the calculation formulas.
To facilitate astrophysical applications, we quantitatively evaluated the percentage increment of the dynamic quantities $L_{\text{peak}}$, $V_f$, $M_f$, and $\alpha_f$ relative to their circular orbit counterparts. This analysis provides a useful measure of the deviations of the dynamic quantities from circular orbits and the impact of eccentricity on the dynamical properties of the system.

In the spin-aligned case, we observe a similarity in the overall behavior of the dynamic quantities compared to the non-spinning scenario. However, the presence of the hangup effect introduces modulations in the relationship between the initial eccentricity and the dynamical quantities, relative to the nonspinning case. In particular, we make a significant discovery in this context. That is, when the spin angular momentum and orbital angular momentum are anti-aligned, we find that by adjusting the initial eccentricity, which is equivalent to modifying the initial tangential momentum, the spin of the final remnant $\alpha_f$ can undergo a transition from positive to negative, passing through the Schwarzschild black hole configuration along the way.
Furthermore, we discover that the percentage of increments of the dynamic quantities with respect to the initial eccentricity in the spin-aligned BBH systems is similar to that observed in the non-spinning case. This finding highlights the consistency in the effects of eccentricity on the dynamics of both spin-aligned and nonspinning BBH systems. 

In the spin-precessing case, we also observe a general similarity in the overall behavior of the dynamic quantities compared to the nonspinning and spin-aligned cases. However, we note distinct characteristics in the recoil velocities, which exhibit larger magnitudes and more intricate curves compared to the previous two scenarios. Through a comprehensive analysis, we conclude that these complex recoil behaviors arise from the combined influence of spin precession and eccentricity.
Furthermore, we find that the percentage increment of the dynamic quantities with respect to the initial eccentricity follows a pattern similar to that observed in the nonspinning and spin-aligned cases. These observations underscore the universality of the effect of eccentricity on the dynamics of BBH systems, regardless of the presence or absence of spin.

All in all, our comprehensive analysis reveals universal behavior in the influence of eccentricity on BBH dynamics. This behavior can be described as follows: Initially, the effect of eccentricity is minimal, resulting in nearly horizontal straight-line trajectories. As eccentricity increases, the dynamic quantities, including peak luminosity $L_{\text{peak}}$, recoil velocity $V_f$, mass $M_f$, and spin $\alpha_f$, exhibit gradual oscillations, reaching peaks or valleys at certain points. As eccentricity further increases, under high eccentricity and head-on collision limits, the dynamic quantities tend to converge towards specific values.
This unified model provides a comprehensive understanding of how the initial eccentricity influences the various dynamic quantities in BBH systems of different mass ratios and spin configurations, encompassing the entire range from low to high eccentricities.

However, it is essential to acknowledge the limitations of our current study. While we have made significant progress, it remains incomplete. For the initial coordinate separation of $11.3M$, although we have a substantial number of data points, the density is still not sufficient to draw definitive conclusions. Similarly, in the case of $24.6M$, the initial eccentricity is not small enough, and the number of data points is limited.
To develop a more comprehensive understanding, it is necessary to investigate other coordinate separations and analyze the unified behavior of the influence of eccentricity on dynamic quantities. Additionally, various factors, including errors arising from insufficient data point density, uncertainties in measured eccentricity, numerical inaccuracies, and the effects of periastron precession, need to be thoroughly addressed. Therefore, further research utilizing eccentric orbital numerical simulations is needed to verify these findings and address these challenges.
Furthermore, the absence of trajectory information in the RIT dataset hinders our ability to fully analyze the dynamic origins of the observed oscillations. Incorporating trajectory information into future studies will be important to gain deeper insight into this phenomenon.
Moreover, the cases of spin alignment and spin precession explored in this study do not cover a sufficiently wide parameter space in terms of spin, initial eccentricity, and mass ratio. The limited number of numerical simulation data points in these cases may restrict the generalizability of the results.

Moving forward, as numerical relativistic simulations of eccentric orbit BBH mergers continue to advance, the influence of eccentricity on dynamics will gradually be revealed. A more practical approach for astrophysical applications would be to develop analytical models that describe the relationship between the dynamic quantities, such as $T_{\text{merger}}$, $L_{\text{peak}}$, $V_f$, $M_f$, and $\alpha_f$, in terms of the initial eccentricity $e_0$. Investigating and constructing such unified models will be the main focus of our future research endeavors.

\begin{acknowledgments}
The authors are very grateful to the RIT collaboration for the numerical simulation of eccentric BBH mergers, and thanks to Yan-Fang Huang, Zhou-Jian Cao, Duan-Yuan Gao, Lin Zhou, Yuan-Yuan Zuo, Jun-Yi Shen, Dong-Jie Liu and Shi-Yan Tian for their helpful discussions. The computation is partially completed in the HPC Platform of Huazhong University of Science and Technology. The language was polished by ChatGPT during the revision of the draft. This work is supported by the National Key R\&D Program of China (2021YFA0718504).
\end{acknowledgments}

\bibliographystyle{apsrev4-2}

\bibliography{ref}
\appendix 
\section{Error Estimate}\label{App:A}
As previously mentioned, numerical errors play a significant role in the occurrence of oscillation phenomena, as they can potentially contribute to such oscillations. While RIT's catalog and metadata do not provide a direct error estimate for each dynamic quantity, related articles offer a rough estimation of the errors. For instance, evaluations of quantities associated with the black hole horizon, such as the final mass and spins of the remnant, indicate errors on the order of 0.1\% through the isolated horizon algorithm. Additionally, radiatively computed quantities, including recoil velocities and peak luminosities, are assessed with a typical error of approximately 5\% \cite{Healy:2022wdn}.
In FIG. \ref{FIG:8}, we present the recoil velocity $V_f$ (panel (a)), peak luminosity $L_{\text{peak}}$ (panel (b)), mass $M_f$ (panel (c)), and spin $\alpha_f$ (panel (d)) with error bars based on the typical error estimate provided by RIT with 5\% for recoil velocity and peak luminosity and 0.1\% for mass and spin. For spin $\alpha_f$, an enlarged view is shown in panel (d) due to its smaller error bar.

By examining panels (a), (b), and (c) in FIG. \ref{FIG:8}, we observe that the last peak or valley for the recoil velocity, peak luminosity, and remnant mass lies completely outside the error bars, indicating that these values are not attributed to numerical errors. However, for penultimate or smaller peaks or valleys, the error bars overlap, preventing us from drawing definitive conclusions. Turning to panels (d) and (e) in FIG. \ref{FIG:8}, we find that the peaks and valleys of the penultimate or smaller spin also fall outside the range of the error bars, suggesting that the oscillation of the remnant spin is a genuine physical phenomenon rather than a result of numerical errors. As elaborated in Section III, Part 3, each physical quantity is derived from the Weyl scalar $\Psi_4$, as denoted by equations (12)-(28). It is important to note that these calculations are interconnected and stem from the same source. Notably, peak luminosity and gravitational wave energy radiation are closely intertwined, with the former serving as the primary contributor to the latter. Consequently, we maintain that if the error estimate for the remnant spin produced by RIT is deemed reasonable, then FIG. \ref{FIG:8} provides evidence to support the conclusion that the observed oscillation is not attributable to numerical errors.  

\begin{figure*}[htbp!]
\centering
\includegraphics[width=15cm,height=15cm]{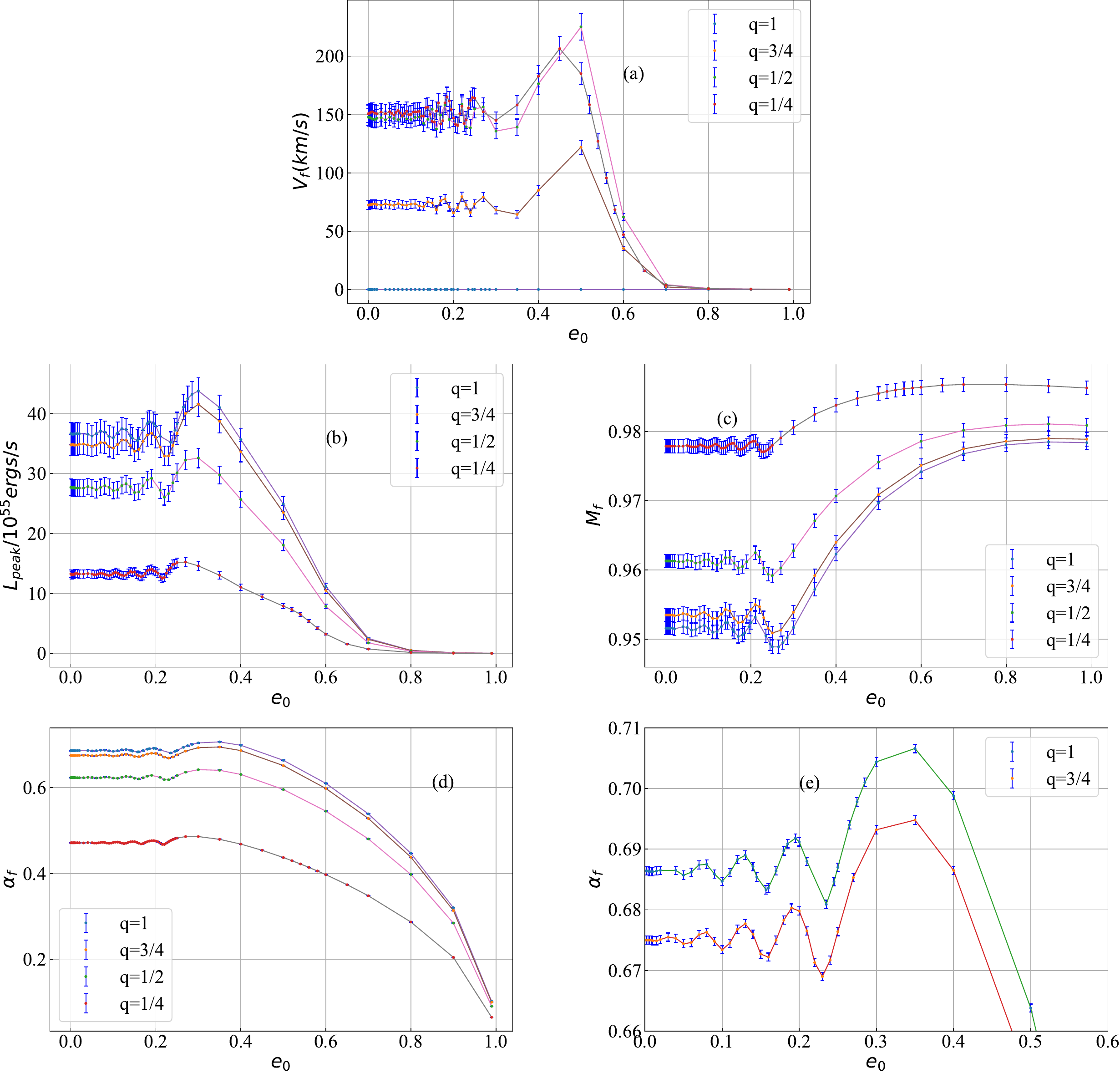}
\caption{\label{FIG:8}Recoil velocity $V_f$ (panel (a)), peak luminosity $L_{\text{peak}}$ (panel (b)), mass $M_f$ (panel (c)), and spin $\alpha_f$ (panel (d)) with error bars based on the typical error estimate provided by RIT with 5\% for recoil velocity and peak luminosity and 0.1\% for mass and spin. For spin $\alpha_f$, an enlarged view is shown in panel (e) due to its smaller error bar.}
\end{figure*}

\begin{figure*}[htbp!]
\centering
\includegraphics[width=15cm,height=5cm]{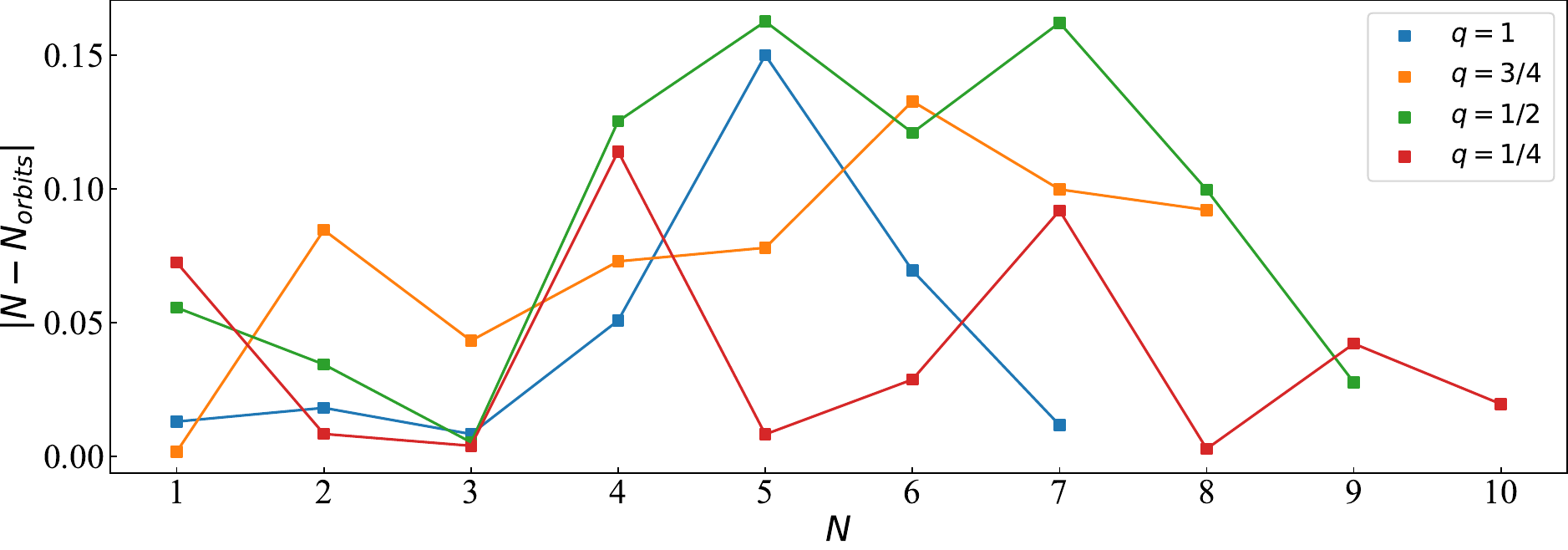}
\caption{\label{FIG:9} Errors of the nearest continuous integer orbital cycle $N_{\text{orbits}}$ compared to the corresponding integer value for the $11.3M$ nonspinning case}
\end{figure*}
\section{Measure eccentricity} \label{App:B}

While the eccentricity estimation method employed by RIT is considered a reliable approximation, it is essential for the integrity of our research to seek a more reasonable approach for estimating the initial eccentricity of the BBH simulation. Similar to Ref. \cite{Sperhake:2007gu} and \cite{Sperhake:2019wwo}, we use the generalized 3PN quasi-Keplerian parameterization to estimate the initial eccentricity. Measuring initial eccentricity can be expressed by Eqs.  (21a) and (25d) in Ref. \cite{Memmesheimer:2004cv}, under ADM and harmonious coordinates. Here, we choose the time eccentricity $e_t$ as the object, but we can also choose other eccentricities such as $e_r$ and $e_{\phi}$. How to choose eccentricity is only a quantitative expression, and it is not necessary to choose $e_t$. All that matters in calculating eccentricity is the initial binding energy $E_b$ and initial angular momentum $L$, and these are available in the metadata of the RIT catalog. What we need to pay attention to is that when the eccentricity is too large, due to the limitations of PN calculation, we may calculate an eccentricity much greater than 1 if the BBH merger is in the plunge phase. In FIGs. \ref{FIG:10} and \ref{FIG:11}, we show the initial eccentricity we obtained under ADM coordinate ($e_\text{ADM,0}$) and harmonic coordinate ($e_\text{harmonic,0}$). Among them, some simulated eccentricities were abandoned by us because they were far greater than 1, and these eccentricities are obtained all in the serious plunge stage. They are located at the tail end of the curve and have no impact on the overall trend of the curve. In FIG. \ref{FIG:12}, we compare the measurement of dynamic quantities by three eccentricities. We observed that in the range of low to medium eccentricity (0-0.4), the eccentricities of ADM coordinate and harmonic coordinate measurement are basically consistent, while in the case of high eccentricity (0.4-1), the two measurement methods deviate. The RIT measurement method ($e_\text{RIT,0}$) deviates from the eccentricity measured by ADM and harmonic coordinates throughout the entire eccentricity range (0-1), and therefore it can only represent an approximate representation. When we need to consider specific initial eccentricity values, we need to mainly refer to the results obtained by PN.

\begin{figure*}[htbp!]
\centering
\includegraphics[width=15cm,height=15cm]{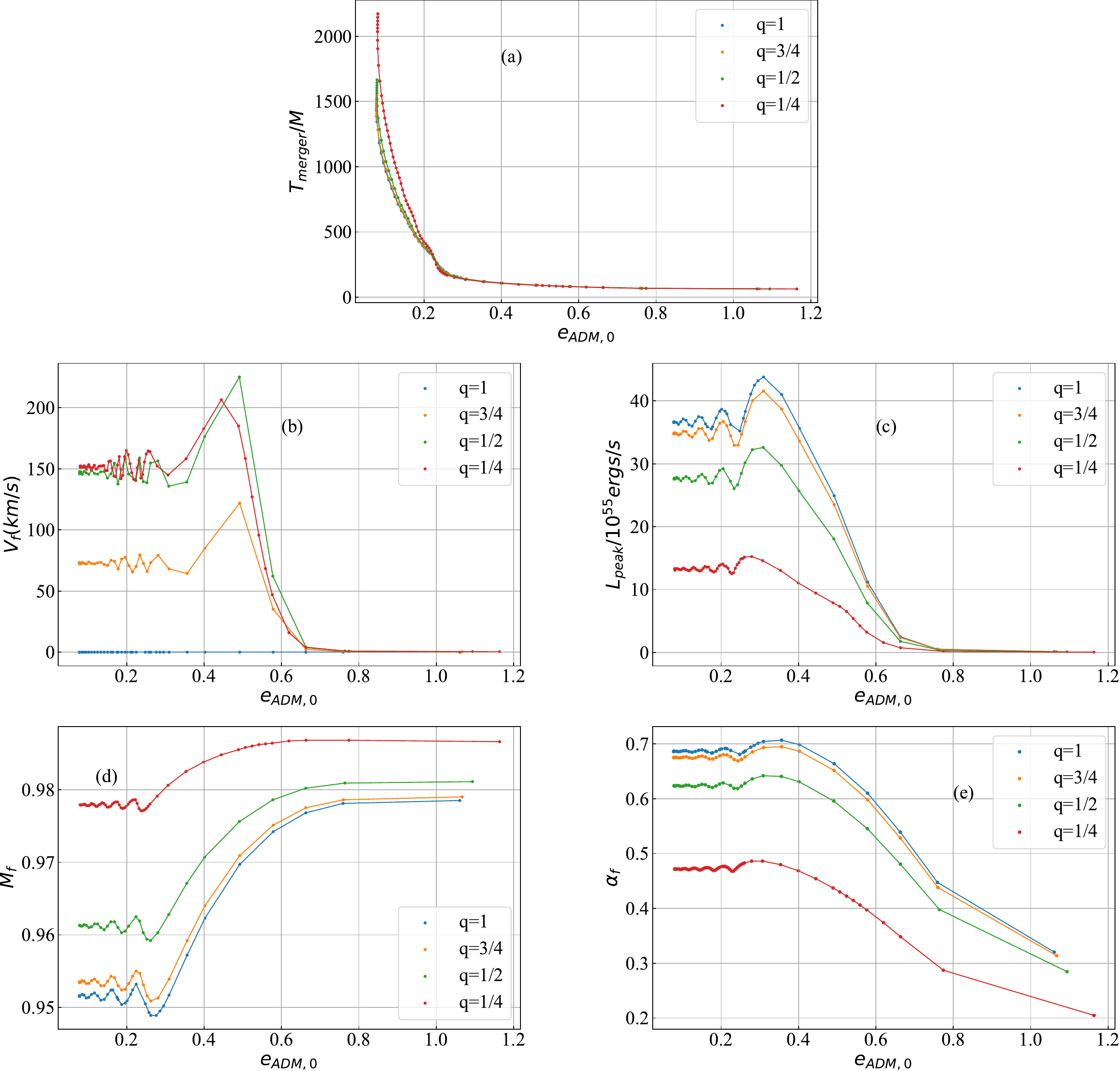}
\caption{\label{FIG:10}Variations of dynamical quantities of the merger time $T_{\text{merger}}$ (panel (a)), peak luminosity $L_{\text{peak}}$ (panel (c)), recoil velocity $V_f$ (panel (b)), mass $M_f$ (panel (d)), and spin $\alpha_f$ (panel (e)) of the merger remnants as a function of the initial eccentricity $e_\text{ADM,0}$ at the initial coordinate separation of $11.3M$ for nonspinning configuration. Initial eccentricities are measured under ADM coordinate of 3PN.}
\end{figure*}

\begin{figure*}[htbp!]
\centering
\includegraphics[width=15cm,height=15cm]{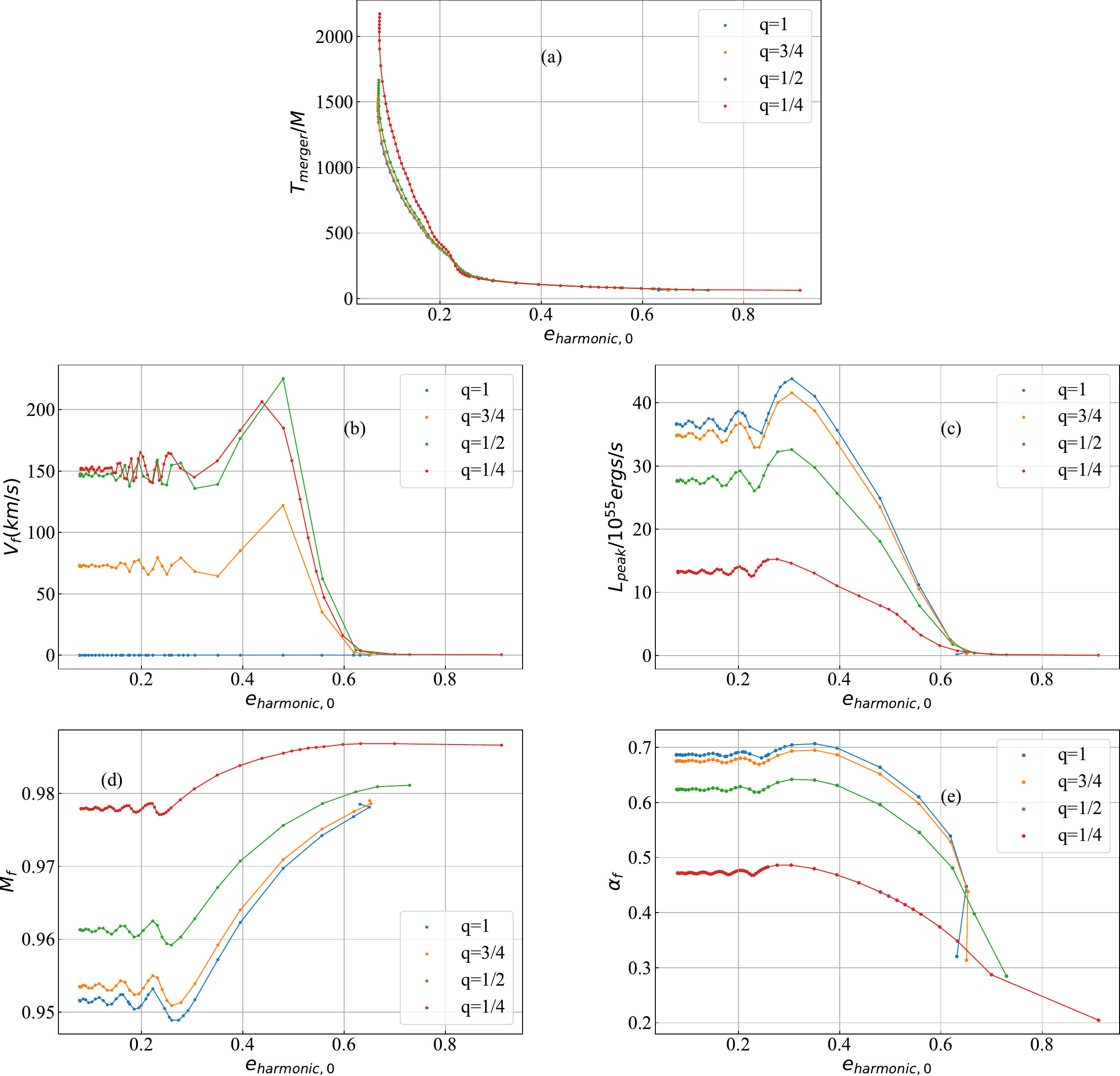}
\caption{\label{FIG:11}Variations of dynamical quantities of the merger time $T_{\text{merger}}$ (panel (a)), peak luminosity $L_{\text{peak}}$ (panel (c)), recoil velocity $V_f$ (panel (b)), mass $M_f$ (panel (d)), and spin $\alpha_f$ (panel (e)) of the merger remnants as a function of the initial eccentricity $e_\text{harmonic,0}$ at the initial coordinate separation of $11.3M$ for nonspinning configuration. Initial eccentricities are measured under harmonic coordinate of 3PN.}
\end{figure*}

\begin{figure*}[htbp!]
\centering
\includegraphics[width=15cm,height=15cm]{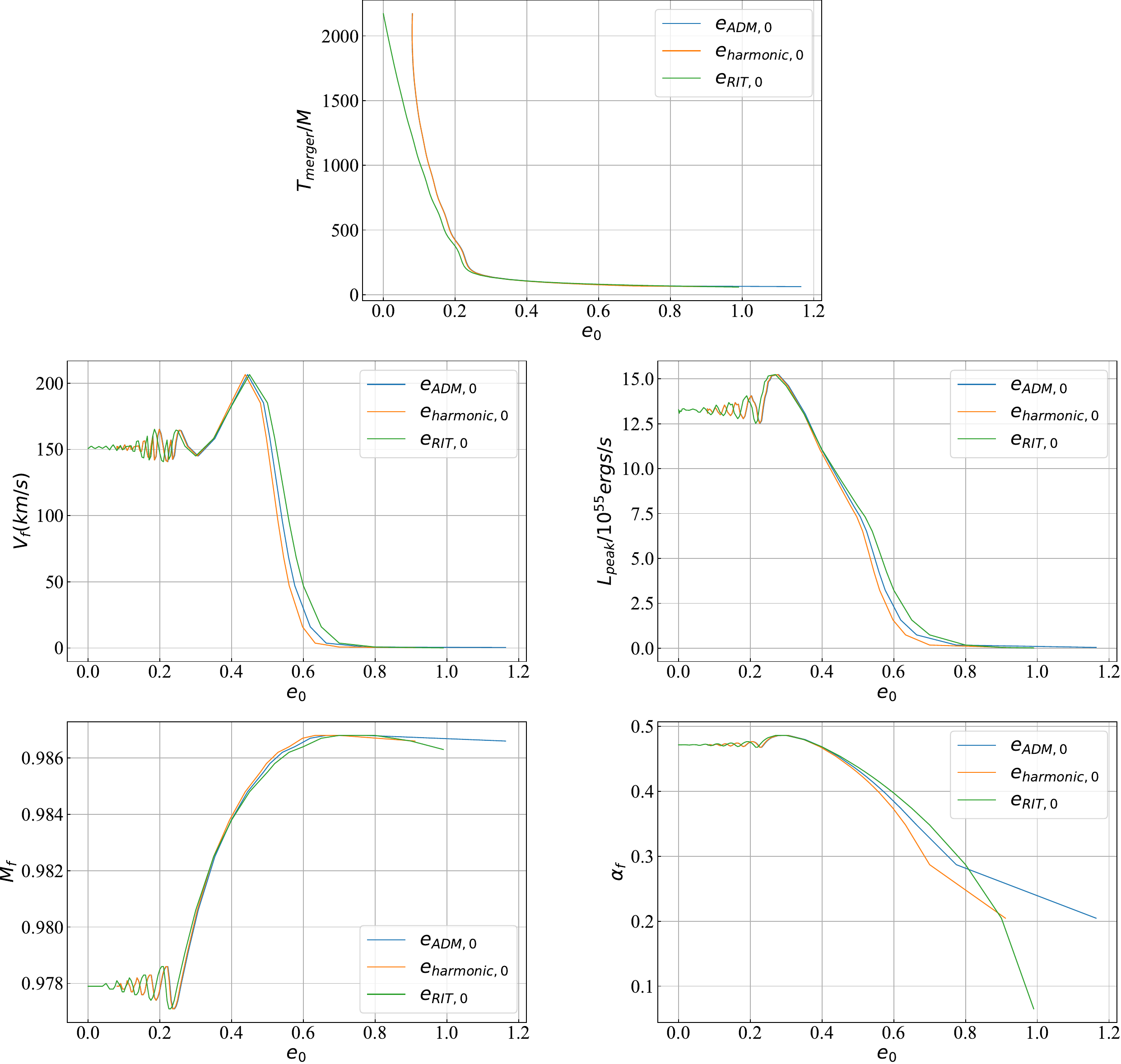}
\caption{\label{FIG:12}Variations of dynamical quantities of the merger time $T_{\text{merger}}$ (panel (a)), peak luminosity $L_{\text{peak}}$ (panel (c)), recoil velocity $V_f$ (panel (b)), mass $M_f$ (panel (d)), and spin $\alpha_f$ (panel (e)) of the merger remnants as a function of the initial eccentricity at the initial coordinate separation of $11.3M$ for nonspinning configuration of mass ratio $q=1/4$. Initial eccentricities are measured through three methods $e_\text{ADM,0}$, $e_\text{harmonic,0}$, $e_\text{RIT,0}$.}
\end{figure*}

\section{Comparison of different initial separations}\label{App:C}
In FIGs. \ref{FIG:13}, \ref{FIG:14} and \ref{FIG:15}, we compare different dynamic quantities under initial coordinate separations of $11.3M$ and $24.6M$ for the nonspinning configuration under three eccentricity measurement methods (ADM, harmonic, RIT). Among them, for the case of $24.6M$, we only list the one with mass ratio $q=1$, because its data points are dense enough and obvious oscillations occur. For the $11.3M$ and $24.6M$ simulations, the sole distinction between them resides in their initial separations. Both simulations entail nonspinning configurations, featuring initial eccentricity values spanning from 0 to 1. Comparing these simulations directly enables us to scrutinize the impact of varying initial separations on the dynamic quantities. In Sec. \ref{sec:III}, we have expounded upon how contrasting two sets of simulations with different initial separations offers valuable insights into the oscillation behavior of dynamic quantities with eccentricity within the strong field regime. As the initial separation increases, the oscillations become discernible at higher initial eccentricities, while also exhibiting more conspicuous patterns. This phenomenon stems directly from the augmented initial separation and the resulting escalated velocity of the binary black hole prior to merger. Through the direct comparison of simulations characterized by distinct initial distances, we effectively capture and elucidate this characteristic.

\begin{figure*}[htbp!]
\centering
\includegraphics[width=15cm,height=15cm]{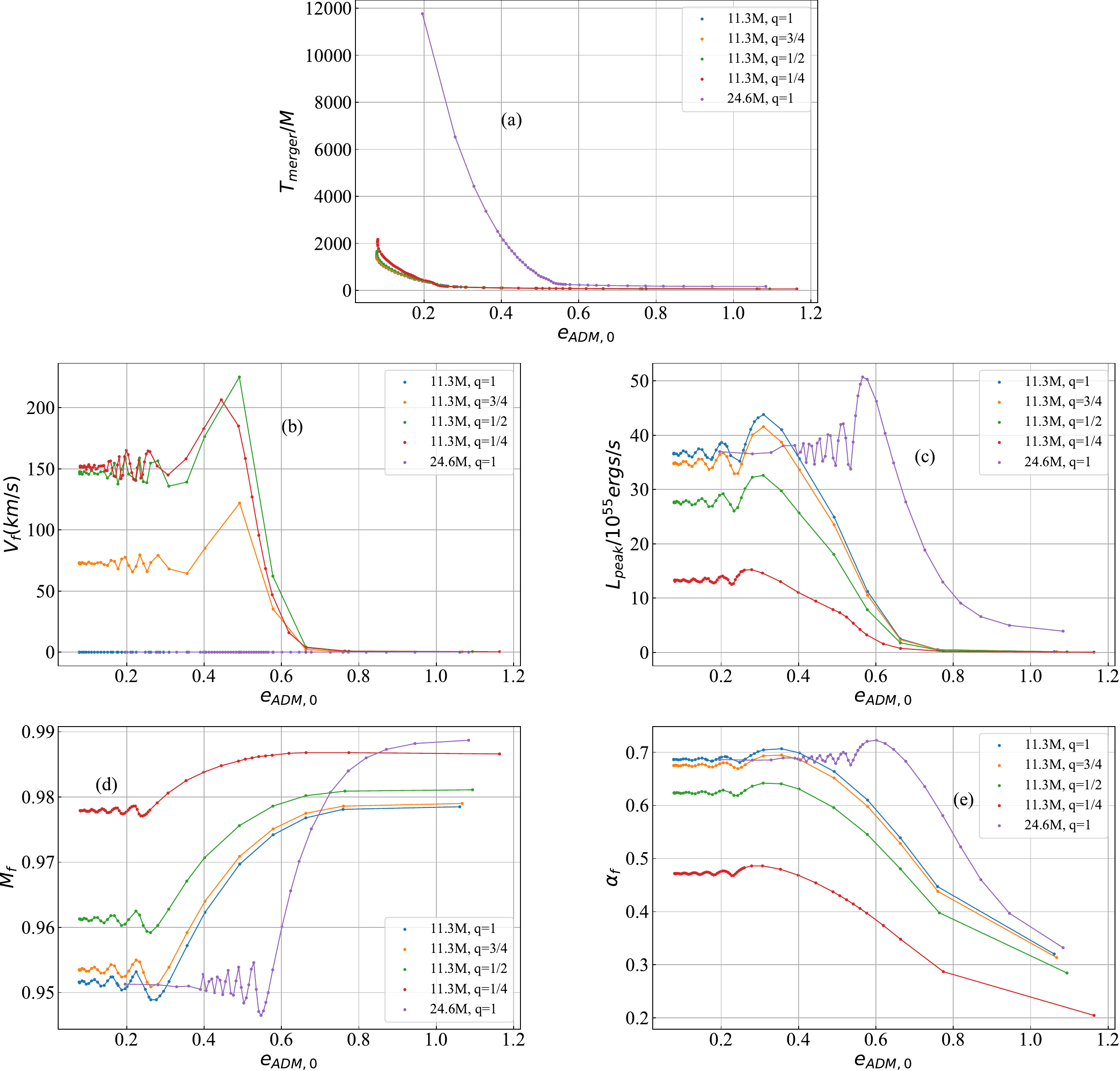}
\caption{\label{FIG:13}
Comparisons of dynamic quantities merger time $T_{\text{merger}}$ (panel (a)), recoil velocity $V_f$ (panel (b)), peak luminosity $L_{\text{peak}}$ (panel (c)), mass $M_f$ (panel (d)), and spin $\alpha_f$ (panel (e)) under initial coordinate separations of $11.3M$ and $24.6M$ (only for mass ratio $q=1$) for the nonspinning configuration under eccentricity measurement methods in ADM coordinates.}
\end{figure*}

\begin{figure*}[htbp!]
\centering
\includegraphics[width=15cm,height=15cm]{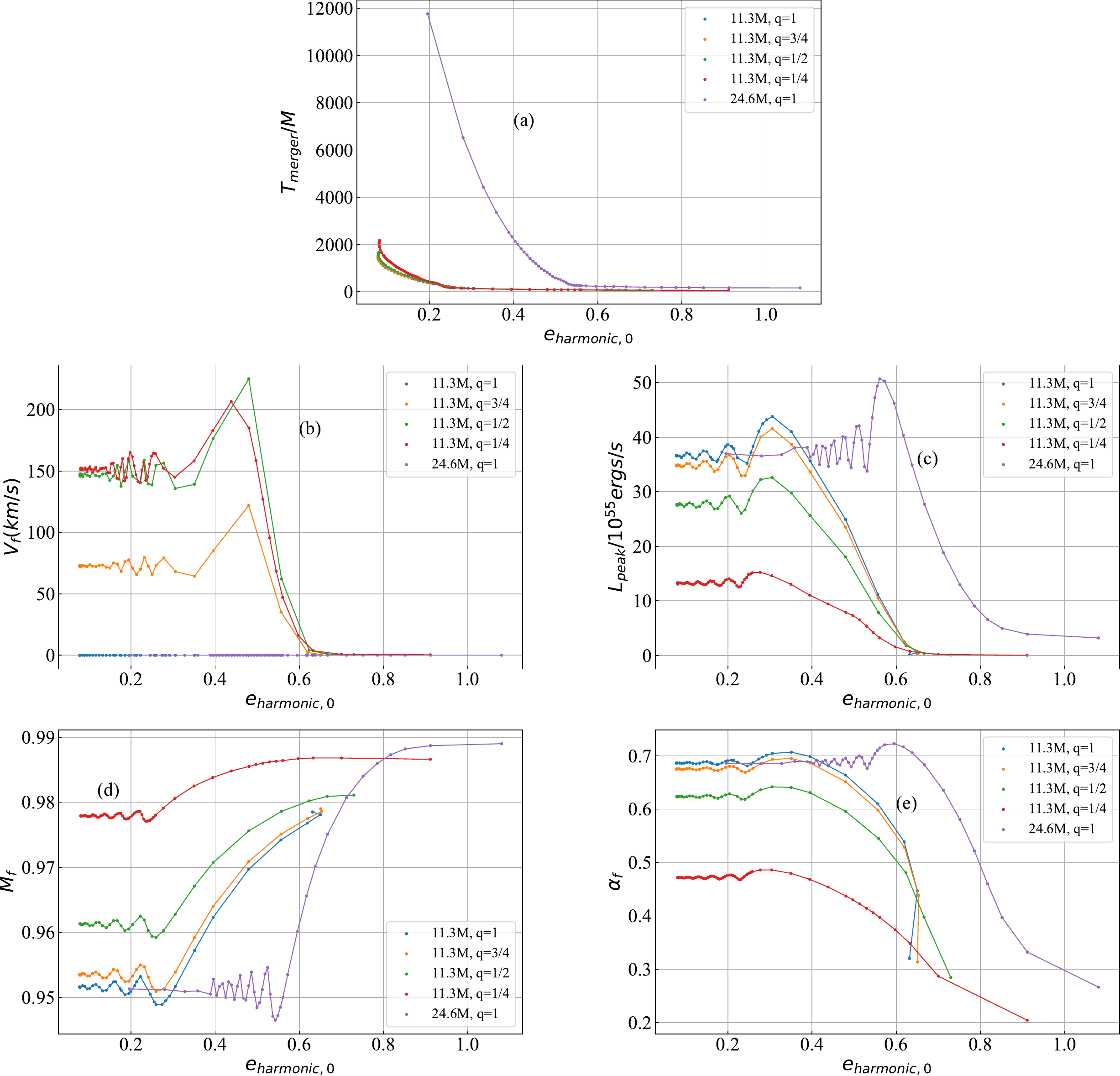}
\caption{\label{FIG:14}Comparisons of dynamic quantities merger time $T_{\text{merger}}$ (panel (a)), recoil velocity $V_f$ (panel (b)), peak luminosity $L_{\text{peak}}$ (panel (c)), mass $M_f$ (panel (d)), and spin $\alpha_f$ (panel (e)) under initial coordinate separations of $11.3M$ and $24.6M$ (only for mass ratio $q=1$) for the nonspinning configuration under eccentricity measurement methods in harmonic coordinates.}
\end{figure*}

\begin{figure*}[htbp!]
\centering
\includegraphics[width=15cm,height=15cm]{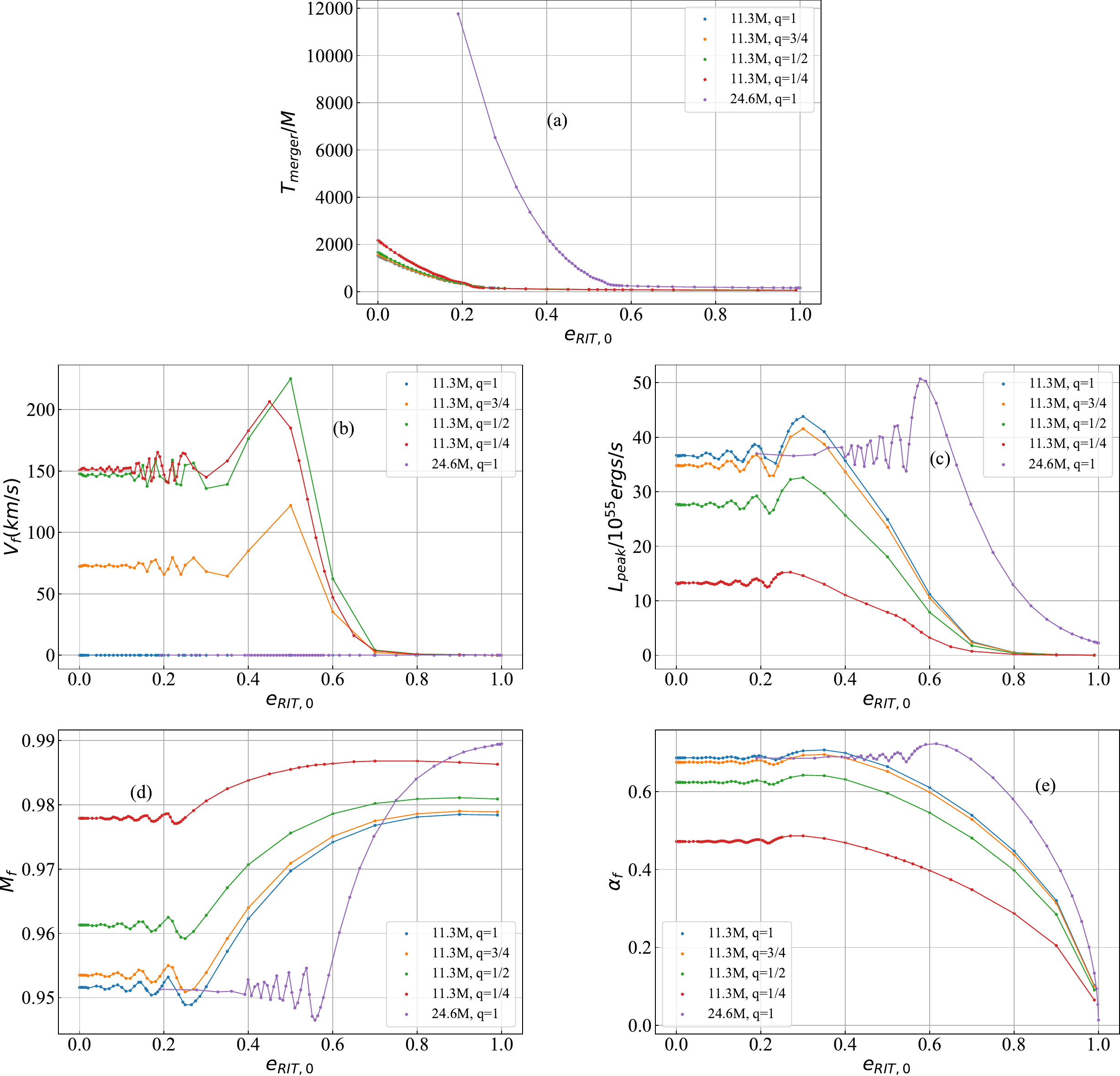}
\caption{\label{FIG:15}Comparisons of dynamic quantities merger time $T_{\text{merger}}$ (panel (a)), recoil velocity $V_f$ (panel (b)), peak luminosity $L_{\text{peak}}$ (panel (c)), mass $M_f$ (panel (d)), and spin $\alpha_f$ (panel (e)) under initial coordinate separations of $11.3M$ and $24.6M$ (only for mass ratio $q=1$) for the nonspinning configuration under eccentricity measurement methods from RIT.}
\end{figure*}

\end{document}